\documentclass[aps,groupedaddress,showpacs]{revtex4}

\usepackage{graphicx}
\DeclareGraphicsExtensions{.ps}

\usepackage{float}

\begin{document}
\preprint{IPP-GEM-2}
\title{Dynamical Alignment in Three Species Tokamak Edge Turbulence}
\author{Bruce D. Scott}
\email[email: ]{bds@ipp.mpg.de}
\homepage[\\ URL: ]{http://www.rzg.mpg.de/~bds/}
\affiliation{Max-Planck-Institut f\"ur Plasmaphysik, 
		Euratom Association,
                D-85748 Garching, Germany}

\date{\today}

\begin{abstract}
Three dimensional computations of self consistent
three species gyrofluid turbulence are carried out for tokamak edge
conditions.  Profiles as well as disturbances in dependent variables are
followed, running the dynamical system to transport equilibrium.  The
third species density shows a significant correlation with that of
the electrons, regardless of initial conditions and drive mechanisms.
For decaying systems the densities evolve toward each other.  Companion
tests with a simple two dimensional drift wave model show this persists
even if the third species is a passively advected test field.
Similarity in the transport character of electrons and the trace
species does not imply that the electrons themselves have a test
particle transport character.
\end{abstract}

\pacs{52.25.Fi,   52.65.Tt,   52.35.Ra,   52.30.Ex}

\maketitle

\def\emskip{\hskip 1 em}
\def\hfb{\hfil\break}
\def\etc{{\it etc.}}
\def\visavis{{\it vis-a-vis}\ }
\def\ie{{\it i.e.}}
\def\eg{{\it e.g.}}
\def\etal{{\it et al}}
\def\ua{u.a.\ }
\def\dh{d.h.\ }
\def\zb{z.B.\ }
\def\bzw{bzw.\ }
\def\usw{usw.\ }

\def\idelta{$i$-delta}


\def\half{ {1\over 2} }
\def\third{ {1\over 3} }
\def\fourth{ {1\over 4} }
\def\tth{ {2\over 3} }
\def\twothirds{ {2\over 3} }
\def\threehalves{ {3\over 2} }
\def\fivehalves{ {5\over 2} }
\def\fivethirds{ {5\over 3} }
\def\sevenhalves{ {7\over 2} }
\def\threeh{\threehalves}
\def\eps{\epsilon}
 
\def\grapprox{\mathop{\lower.5ex \hbox{$\buildrel{\fivesy >}\over{\fivesy\sim}$}} \nolimits}
\def\lsapprox{\mathop{\lower.5ex \hbox{$\buildrel{\fivesy <}\over{\fivesy\sim}$}} \nolimits}
\def\grls{\mathop{\lower.5ex \hbox{$\buildrel{\fivesy >}\over{\fivesy <}$}} \nolimits}

\def\vec#1{{\bf #1}}
\def\tsr#1{{\secfnt #1}}
\def\avg#1{\left\langle #1 \right\rangle}
\def\abs#1{\left\vert #1 \right\vert}
\def\prf#1{\overline{#1}}

\def\max{{}_{{\rm max}}}
\def\min{{}_{{\rm min}}}

\def\minus{\mathop{\hbox{--}}\nolimits}

\def\re{\mathop{\rm Re}\nolimits}
\def\im{\mathop{\rm Im}\nolimits}
\def\sech{\mathop{\rm sech}\nolimits}
\def\diag{\mathop{\rm diag}\nolimits}
\def\Max{\mathop{\rm Max}\nolimits}
\def\Min{\mathop{\rm Min}\nolimits}
\def\nint{\mathop{\rm NINT}\nolimits}
\def\mod{\mathop{\rm mod}\nolimits}
\def\det{\mathop{\rm det}\nolimits}
\def\Tr{\mathop{\rm Tr}\nolimits}
\def\sign{\mathop{\rm sign}\nolimits}

\def\LBR{\left\lbrace}
\def\RBR{\right\rbrace}
\def\LB{\left\lbrack}
\def\RB{\right\rbrack}
\def\LP{\left (}
\def\RP{\right )}
\def\qq{\qquad\qquad}
\def\qqq{\qquad\qquad\qquad}
\def\Det#1{\left\vert\matrix{#1}\right\vert}

\def\pt{\partial}

\def\pzz#1{{\partial #1\over\partial z}}
\def\pxx#1{{\partial #1\over\partial x}}
\def\pyy#1{{\partial #1\over\partial y}}
\def\pww#1{{\partial #1\over\partial w}}
\def\pss#1{{\partial #1\over\partial s}}
\def\prr#1{{\partial #1\over\partial r}}
\def\prhrh#1{{\partial #1\over\partial \rho}}
\def\pthth#1{{\partial #1\over\partial \theta}}
\def\pchch#1{{\partial #1\over\partial \chi}}
\def\ppsps#1{{\partial #1\over\partial \psi}}
\def\pzeze#1{{\partial #1\over\partial \zeta}}
\def\pphph#1{{\partial #1\over\partial \phi}}
\def\ptt#1{{\partial #1\over\partial t}}
\def\pVV#1{{\partial #1\over\partial V}}
\def\phh#1{{\partial #1\over\partial \theta}}
\def\pvhvh#1{{\partial #1\over\partial \vartheta}}
\def\pxixi#1{{\partial #1\over\partial \xi}}
\def\dtt#1{{d #1\over dt}}
\def\dss#1{{d #1\over ds}}
\def\drr#1{{d #1\over dr}}
\def\pprr#1{{\partial^2 #1\over\partial r^2}}
\def\pprhrh#1{{\partial^2 #1\over\partial \rho^2}}
\def\ppss#1{{\partial^2 #1\over\partial s^2}}
\def\ppxx#1{{\partial^2 #1\over\partial x^2}}
\def\ppxy#1{{\partial^2 #1\over\partial x\partial y}}
\def\ppyy#1{{\partial^2 #1\over\partial y^2}}
\def\ppys#1{{\partial^2 #1\over\partial y\partial s}}
\def\ppzz#1{{\partial^2 #1\over\partial z^2}}
\def\pptt#1{{\partial^2 #1\over\partial t^2}}
\def\ppVV#1{{\partial^2 #1\over\partial V^2}}
\def\ppphph#1{{\partial^2 #1\over\partial \phi^2}}
\def\ppthth#1{{\partial^2 #1\over\partial \theta^2}}
\def\pphh#1{{\partial^2 #1\over\partial \theta^2}}
\def\ppvhvh#1{{\partial^2 #1\over\partial \vartheta^2}}
\def\ppxixi#1{{\partial^2 #1\over\partial \xi^2}}
\def\ppzeze#1{{\partial^2 #1\over\partial \zeta^2}}
\def\pphze#1{{\partial^2 #1\over\partial\theta\partial\zeta}}
\def\ppz#1{\partial #1/\partial z}
\def\ppx#1{\partial #1/\partial x}
\def\ppy#1{\partial #1/\partial y}
\def\ppw#1{\partial #1/\partial w}
\def\ppr#1{\partial #1/\partial r}
\def\pprh#1{\partial #1/\partial \rho}
\def\pps#1{\partial #1/\partial s}
\def\ppt#1{\partial #1/\partial t}
\def\ppV#1{\partial #1/\partial V}
\def\pph#1{\partial #1/\partial \theta}
\def\ppvh#1{\partial #1/\partial \vartheta}
\def\ppxi#1{\partial #1/\partial \xi}

\def\ddt#1{d #1/dt}
\def\pppz#1{\partial^2 #1/\partial z^2}
\def\pppx#1{\partial^2 #1/\partial x^2}
\def\pppy#1{\partial^2 #1/\partial y^2}
\def\pppr#1{\partial^2 #1/\partial r^2}
\def\ppprh#1{\partial^2 #1/\partial \rho^2}
\def\ppps#1{\partial^2 #1/\partial s^2}
\def\pppt#1{\partial^2 #1/\partial t^2}
\def\pppV#1{\partial^2 #1/\partial V^2}
\def\ppph#1{\partial^2 #1/\partial \theta^2}
\def\pppvh#1{\partial^2 #1/\partial \vartheta^2}
\def\pppxi#1{\partial^2 #1/\partial \xi^2}
\def\dddt#1{d^2 #1/dt^2}

\def\grad{\nabla}
\def\cross{{\bf \times}}
\def\div{\grad\cdot}
\def\divp{\grad_\perp\cdot}
\def\divpl{\grad_\parallel\cdot}
\def\curl{\grad\cross}
\def\dpl{\grad_\parallel}
\def\ddpl{\grad_\parallel^2}
\def\dpp{\grad_\perp}
\def\ddpp{\grad_\perp^2}
\def\delsq{\grad^2}
\def\delamb{ \mathchar"0274\hskip -.665em\mathchar"0275 }
\let\delam=\delamb
\def\lapl{\grad^2}
\def\lapldef{\ddpp=(\pt^2/\pt x^2)+K^2(\pt^2/\pt y^2)}

\def\pwww#1{{\partial #1\over\partial \vec w}}
\def\pwwpl#1{{\partial #1\over\partial w_\parallel}}
 
\def\pvv#1#2{{\partial #2\over\partial v_{#1}}}
\def\ppv#1#2{{\partial #2/\partial v_{#1}}}
\def\pvvv#1{{\partial #1\over\partial \vec v}}
\def\pvvp#1#2{{\partial #2\over\partial v'_{#1}}}
\def\ppvp#1#2{{\partial #2/\partial v'_{#1}}}
\def\pvvvp#1{{\partial #1\over\partial \vec v'}}
\def\pvvpl#1{{\partial #1\over\partial v_\parallel}}
 
\def\xunit{\vec{\hat x}}
\def\yunit{\vec{\hat y}}
\def\zunit{\vec{\hat z}}
\def\sunit{\vec{\hat s}}
\def\bunit{\vec{b}}
\def\eunit{\vec{\hat e}}
\def\nunit{\vec{\hat n}}
\def\dt{\Delta t}
\def\becomes{\leftarrow}
\def\from{\leftarrow}
\def\to{\rightarrow}
\def\fromto{\leftrightarrow}
\def\implies{\,\,\,\Longrightarrow\,\,\,}
\def\dotdot{\!:\!}

\def\meters{\,{\rm m}}
\def\invm{\,{\rm m}^{-3}}
\def\invsec{\,{\rm sec}^{-1}}
\def\cm{\,{\rm cm}}
\def\km{\,{\rm km}}
\def\invcc{\,{\rm cm}^{-3}}
\def\invcm{\,{\rm cm}^{-1}}
\def\invmm{\,{\rm mm}^{-1}}
\def\mm{\,{\rm mm}}
\def\Vcm{\,{\rm V/cm}}
\def\Acm{\,{\rm A/cm^2}}
\def\kA{\,{\rm kA}}
\def\MA{\,{\rm MA}}
\def\degk{\,{\rm K}}
\def\ergs{\,{\rm erg}}
\def\eV{\,{\rm eV}}
\def\keV{\,{\rm keV}}
\def\MeV{\,{\rm MeV}}
\def\GeV{\,{\rm GeV}}
\def\kG{\,{\rm kG}}
\def\tesla{\,{\rm T}}
\def\kW{\,{\rm kW}}
\def\MW{\,{\rm MW}}
\def\MWsqm{\,{\rm MW/m^2}}
\def\Wsqm{\,{\rm W/m^2}}
\def\radsec{\,{\rm rad/sec}}
\def\Hz{\,{\rm Hz}}
\def\kHz{\,{\rm kHz}}
\def\MHz{\,{\rm MHz}}
\def\mpersec{\,{\rm m}/{\rm sec}}
\def\msqsec{\,{\rm m^2}/{\rm sec}}
\def\cmsec{\,{\rm cm}/{\rm sec}}
\def\kmsec{\,{\rm km}/{\rm sec}}
\def\ccpersec{\,{\rm cm}^3/{\rm sec}}
\def\minutes{\,{\rm min}}
\def\yr{\,{\rm yr}}
\def\hr{\,{\rm hr}}
\def\Bar{\,{\rm bar}}
\def\sec{\,{\rm sec}}
\def\msec{\,{\rm msec}}
\def\usec{\,\mu{\rm sec}}
 
\def\bdel{\vec b\cdot\grad}
\def\Bdel{\vec B\cdot\grad}
\def\Jdel{\vec J\cdot\grad}
\def\bdot{\vec B\cdot}
\def\Bdot{\vec B\cdot}
\def\exb{\vec E\cross\vec B}
\def\jxb{\vec J\cross\vec B}
\def\uxb{\vec u\cross\vec B}
\def\vxb{\vec v\cross\vec B}
\def\wxb{\vec w\cross\vec B}
\def\ucxb{{\vec u\over c}\cross\vec B}
\def\vcxb{{\vec v\over c}\cross\vec B}
\def\wcxb{{\vec w\over c}\cross\vec B}
\def\jcxb{{\vec J\cross\vec B\over c}}

\def\vexb{\vec v_E}
\def\vpol{\vec v_p}
\def\upol{\vec u_p}
\def\vstar{\vec v_*}
\def\ustar{\vec u_*}
\def\Jstar{\vec J_*}
\def\Jpol{\vec J_p}
\def\vgradb{\vec v_{\grad B}}
\def\qstar{\vec q_\wedge}
\def\qestar{\vec q_e{}_\wedge}
\def\qistar{\vec q_i{}_\wedge}
\def\pistar{\vec\Pi_*}
\def\vR{\vec v_R}
\def\vdl{\vec v\cdot\grad}
\def\vdel{\vec v\cdot\grad}
\def\vedl{\vexb\cdot\grad}
\def\udl{\vec u\cdot\grad}
\def\udel{\vec u\cdot\grad}
\def\uidl{\vec u_i\cdot\grad}
\def\uidel{\vec u_i\cdot\grad}
\def\wdel{\vec w\cdot\grad}
\def\dedt#1{d_E #1/dt}
\def\dett#1{{d_E #1\over dt}}
\def\jpp{J_\perp}
\def\jperp{\vec\jpp}
\def\qpp{q_\perp}
\def\qperp{\vec\qpp}
\def\upp{u_\perp}
\def\uperp{\vec\upp}
\def\wpl{w_\parallel}
\def\wpp{w_\perp}
\def\wperp{\vec\wpp}
\def\vpp{v_\perp}
\def\vperp{\vec\vpp}
\def\lnb{\log B}
 
\def\rms{_{rms}}
 
\def\Jpl{J_\parallel}
\def\jpl{J_\parallel}
\def\Jpp{J_\perp}
\def\jpp{J_\perp}
\def\Jperp{\vec\Jpp}
\def\Bperp{\vec B_\perp}
\def\Apl{A_\parallel}
\def\apl{A_\parallel}
\def\App{A_\perp}
\def\app{A_\perp}
\def\Aperp{\vec\App}
\def\Epl{E_\parallel}
\def\epl{E_\parallel}
\def\Epp{E_\perp}
\def\epp{E_\perp}
\def\Eperp{\vec\Epp}
\def\upl{u_\parallel}
\def\vpl{v_\parallel}
\def\Upl{U_\parallel}
\def\vor{\grad_\perp^2\phi}
\def\kpl{k_\parallel}
\def\kkpl{k_\parallel^2}
\def\kpp{k_\perp}
\def\kperp{\vec\kpp}
\def\kkpp{k_\perp^2}
\def\xpl{{x_\parallel}}
\def\xpp{x_\perp}
\def\DD{\Delta_D}
\def\Dpl{D_\parallel}
\def\Dpp{\Delta_\perp}
\def\Depl{D_e{}_\parallel}
\def\Dipl{D_i{}_\parallel}
\def\Rpl{R_\parallel}
\def\qpl{q_\parallel}
\def\qepl{q_e{}_\parallel}
\def\qipl{q_i{}_\parallel}
\def\mupl{\mu_\parallel}
\def\mupp{\mu_\perp}
\def\nuei{\nu_{ei}}
\def\nuee{\nu_{ee}}
\def\nuii{\nu_{ii}}
\def\wpe{\omega_{pe}}
\def\wpi{\omega_{pi}}
\def\nudamp{\nu_d}
\def\zeff{Z_{\!e\!f\!f}}
\def\lmfp{\lambda_{\!m\!f\!p}}
\def\ws{{\omega_*}}
\def\wsi{{\omega_{*i}}}
\def\wn{\omega_n}
\def\wt{\omega_t}
\def\wi{\omega_i}
\def\wT{\omega_T}
\def\wp{\omega_p}
\def\wc{{\omega_c}}
\def\kappacv{{\cal K}}
\def\wcv{{\omega_B}}
\def\etai{\eta_i}
\def\taui{\tau_i}
\def\rs{\rho_s}
\def\ld{\lambda_D}
\def\Lpl{L_\parallel}
\def\Lpp{L_\perp}
\def\lcorpl{\lambda_\parallel}
\def\lcorpp{\lambda_\perp}
\def\rch{\rho_{ch}}
\def\npl{\eta_\parallel}
\def\etapl{\eta_\parallel}
\def\ald{a_L}
\def\alde{a_{Le}}
\def\aldi{a_{Li}}
\def\npp{\eta_\perp}
\def\etapp{\eta_\perp}
\def\kappapl{\kappa_\parallel}
\def\dprime{\Delta'}
\def\sk{{}_{\vec k}}
\def\sky{{}_{k_y}}
\def\gk{\gamma_k}
\def\vk{\vfl_k}
\def\nk{\nfl_k}
\def\tk{\tfl_k}
\def\dk{\Delta k}
\def\gd{\gamma_0}
\def\mwn{\Delta_n}
\def\mwh{\Delta_h}
\def\gamT{\Gamma_T}
\def\gamn{\Gamma_n}
\def\gamt{\Gamma_t}
\def\gami{\Gamma_i}
\def\gamc{\Gamma_c}
\def\gamk{\Gamma_k}
\def\gams{\Gamma_s}
\def\gaml{\Gamma_l}
\def\gamr{\Gamma_r}
 
\def\ptb{\widetilde}
\def\psifl{\widetilde\psi}
\def\phifl{\widetilde\phi}
\def\ffl{\widetilde f}
\def\fe{f_e}
\def\fefl{\widetilde f_e}
\def\fifl{\widetilde f_i}
\def\nfl{\widetilde n}
\def\hfl{\widetilde h}
\def\tfl{\widetilde T}
\def\nefl{\widetilde n_e}
\def\nifl{\widetilde n_i}
\def\tefl{\widetilde T_e}
\def\tifl{\widetilde T_i}
\def\pfl{\widetilde p}
\def\pefl{\widetilde p_e}
\def\pifl{\widetilde p_i}
\def\hefl{\widetilde h_e}
\def\vx{\widetilde v_x}
\def\vfl{\widetilde v}
\def\vefl{\widetilde \vexb}
\def\vxfl{\widetilde v_x}
\def\vyfl{\widetilde v_y}
\def\vrfl{\widetilde v_r}
\def\vppfl{\widetilde v_\perp}
\def\vplfl{\widetilde \vpl}
\def\Bfl{\widetilde \vec B}
\def\Bflpp{\widetilde B_\perp}
\def\Aplfl{\widetilde A_\parallel}
\def\Appfl{\widetilde A_\perp}
\def\Aperpfl{\widetilde {\vec A}_\perp}
\def\ufl{\widetilde u_\parallel}
\def\vorfl{\grad_\perp^2\phifl}
\def\jfl{\widetilde J_\parallel}
\def\qfl{\widetilde q_\parallel}
\def\qefl{\widetilde q_e{}_\parallel}
\def\qifl{\widetilde q_i{}_\parallel}
\def\jppfl{\widetilde J_\perp}
\def\jperpfl{\widetilde {\vec J}_\perp}
\def\Afl{\ptb A_\parallel}
\def\Jfl{\ptb J_\parallel}
\def\efl{\widetilde E_\parallel}
\def\Efl{\widetilde E_\parallel}
\def\Eppfl{\widetilde E_\perp}
\def\Eperpfl{\widetilde {\vec E}_\perp}
\def\etafl{\widetilde\eta}
\def\isatfl{\widetilde I_{{\rm sat}}}
\def\phiflfl{\widetilde\phi_{{\rm fl}}}
 
\def\teplfl{\widetilde T_e{}_\parallel}
\def\teppfl{\widetilde T_e{}_\perp}
\def\qeplfl{\widetilde q_e{}_\parallel}
\def\qeppfl{\widetilde q_e{}_\perp}
\def\tiplfl{\widetilde T_i{}_\parallel}
\def\tippfl{\widetilde T_i{}_\perp}
\def\qiplfl{\widetilde q_i{}_\parallel}
\def\qippfl{\widetilde q_i{}_\perp}

\def\tepl{ T_e{}_\parallel}
\def\tepp{ T_e{}_\perp}
\def\qepl{ q_e{}_\parallel}
\def\qepp{ q_e{}_\perp}
\def\tipl{ T_i{}_\parallel}
\def\tipp{ T_i{}_\perp}
\def\qipl{ q_i{}_\parallel}
\def\qipp{ q_i{}_\perp}

\def\peplfl{\widetilde p_e{}_\parallel}
\def\peppfl{\widetilde p_e{}_\perp}
\def\piplfl{\widetilde p_i{}_\parallel}
\def\pippfl{\widetilde p_i{}_\perp}

\def\pepl{ p_e{}_\parallel}
\def\pepp{ p_e{}_\perp}
\def\pipl{ p_i{}_\parallel}
\def\pipp{ p_i{}_\perp}


\def\phinn{ {e\phifl\over T} }
\def\nnn{ {\nfl\over n} }
\def\tnn{ {\tfl\over T} }
\def\unn{ {\ufl\over c_s} }
\def\vornn{ \rho_s^2\ddpp\phinn }
\def\jnn{ {\jfl\over ne} }
\def\qnn{ {\qfl\over nT} }
\def\psinn{ {\psifl\over B\rho_s} }

\def\ahat{\hat\alpha}
\def\ehat{\hat\eta}
\def\khat{\hat\kappa}
\def\shat{\hat s}
\def\bhat{\hat\beta}
\def\muhat{\hat\mu}
\def\epss{\hat\epsilon}
\def\bigpoint#1{
    \par\bigskip
    {\baselineskip=\normalbaselineskip
    \parindent=0 pt
    {\hfill\vbox{ #1  }\hfill}}
    \par\bigskip
    }
 
\def\jfm#1{{\it J. Fluid. Mech.} {\secfnt #1}}
\def\prl#1{{\it Phys. Rev. Lett.} {\secfnt #1}}
\def\physletta#1{{\it Phys. Lett. A} {\secfnt #1}}
\def\physlettb#1{{\it Phys. Lett. B} {\secfnt #1}}
\def\pf#1{{\it Phys. Fluids} {\secfnt #1}}
\def\pfa#1{{\it Phys. Fluids A} {\secfnt #1}}
\def\pfb#1{{\it Phys. Fluids B} {\secfnt #1}}
\def\physp#1{{\it Phys. Plasmas} {\secfnt #1}}
\def\nf#1{{\it Nucl. Fusion} {\secfnt #1}}
\def\njp#1{{\it New J. Phys.} {\secfnt #1}}
\def\cpp#1{{\it Contrib. Plasma Phys.} {\secfnt #1}}
\def\ppcf#1{{\it Plasma Phys. Contr. Fusion} {\secfnt #1}}
\def\plasphys#1{{\it Plasma Phys.} {\secfnt #1}}
\def\revpp#1{{\it Rev. Plasma Phys.} {\secfnt #1}}
\def\iaea#1#2{in {\it Plasma Physics and Controlled Nuclear Fusion
    Research #1}, Proceedings of the #2th International Conference}
\def\EPS#1#2#3{in {\it Proceedings of the
{#1}th European Conference on Controlled Fusion and Plasma Physics,
{#2}, {#3}} (European Physical Society, {#2}, {#3})}
\def\jcp#1{{\it J. Comput. Phys.} {\secfnt #1}}
\def\jetp#1{{\it Sov. Phys. JETP} {\secfnt #1}}
\def\sovjpp#1{{\it Sov. J. Plasma Phys.} {\secfnt #1}}
\def\jnm#1{{\it J. Nucl. Mat.} {\secfnt #1}}
\def\rsi#1{{\it Rev. Sci. Inst.} {\secfnt #1}}
\def\adv#1{{\it Adv. Phys.} {\secfnt #1}}
\def\apjl#1{{\it Astrophys. J. Lett.} {\secfnt #1}}
\def\apj#1{{\it Astrophys. J.} {\secfnt #1}}
\def\aa#1{{\it Astron. Astrophys.} {\secfnt #1}}
\def\vol#1{\ {\secfnt #1}}

\def\figdecaya{1}

\def\skp{_{\kperp}}
\def\ppb#1{\pt #1/\pt b}
\def\pbb#1{{\pt #1\over\pt b}}
\def\phigfl{\phifl_G}
\def\vorfl{\ptb\Omega}

\def\vpl{\nu_\parallel}
\def\vpp{\nu_\perp}

\def\bb{\vec B}
\def\pyyk#1{{\pt #1\over\pt y_k}}
\def\ppyyk#1{{\pt^2 #1\over\pt y_k^2}}
\def\vex{\ptb v_E^x}
\def\bbx{\ptb b^x}
\def\vey{\ptb v_E^y}
\def\bby{\ptb b^y}
\def\ffl{\ptb f}

\def\wz{\omega_z}

\def\nefl{\widetilde n_e}
\def\teplfl{\widetilde T_e{}_\parallel}
\def\teppfl{\widetilde T_e{}_\perp}
\def\qeplfl{\widetilde q_e{}_\parallel}
\def\qeppfl{\widetilde q_e{}_\perp}
\def\nifl{\widetilde n_i}
\def\Nfl{\widetilde N}
\def\tiplfl{\widetilde T_i{}_\parallel}
\def\tippfl{\widetilde T_i{}_\perp}
\def\qiplfl{\widetilde q_i{}_\parallel}
\def\qippfl{\widetilde q_i{}_\perp}
\def\nzfl{\widetilde n_z}
\def\tzfl{\widetilde T_z}
\def\uzfl{\widetilde u_z{}_\parallel}
\def\qzfl{\widetilde q_z{}_\parallel}
\def\tzplfl{\widetilde T_z{}_\parallel}
\def\tzppfl{\widetilde T_z{}_\perp}
\def\qzplfl{\widetilde q_z{}_\parallel}
\def\qzppfl{\widetilde q_z{}_\perp}

\def\ntfl{\widetilde n_t}
\def\ttfl{\widetilde T_t}
\def\utfl{\widetilde u_t{}_\parallel}
\def\qtfl{\widetilde q_t{}_\parallel}
\def\ttplfl{\widetilde T_t{}_\parallel}
\def\ttppfl{\widetilde T_t{}_\perp}
\def\qtplfl{\widetilde q_t{}_\parallel}
\def\qtppfl{\widetilde q_t{}_\perp}

\def\tepl{ T_e{}_\parallel}
\def\tepp{ T_e{}_\perp}
\def\qepl{ q_e{}_\parallel}
\def\qepp{ q_e{}_\perp}
\def\tipl{ T_i{}_\parallel}
\def\tipp{ T_i{}_\perp}
\def\qipl{ q_i{}_\parallel}
\def\qipp{ q_i{}_\perp}
\def\tzpl{ T_z{}_\parallel}
\def\tzpp{ T_z{}_\perp}
\def\qzpl{ q_z{}_\parallel}
\def\qzpp{ q_z{}_\perp}

\def\dztt#1{{d_z #1\over dt}}
\def\kkpp{k_\perp^2}

\def\uexb{\vec u_E}
\def\uedl{\uexb\cdot\grad}

\def\fhat{\vec{\widehat F}}
\def\fdel{\fhat\cdot\grad}

\section{Introduction --- Trace Species Transport
\label{sec:intro}}

It is an interesting question whether the electrons and ions in a
magnetically confined plasma are transported in the same way test
particles in a randomly turbulent velocity field would be
\cite{horton,manfredi,beyer,basu}.
One might consider to investigate this by placing test particles or a
trace species into an otherwise self consistent turbulence computation
or experimentally by measuring trace species transport and comparing to
the transport of the electrons \cite{traceexpt}.
Not only the transport coefficients are interesting, but also the basic
character, namely, whether or not the transport is a random walk
process \cite{florin,volkertest,annibaldi,annibaldi2}. 
It might be tempting to conclude that if the transport of the trace
species were the same as that of the electrons, then the electrons
themselves should be transported essentially randomly.

It is useful to provide a falsification constraint on this logic.
Gradient driven transport in a magnetised plasma is usually very
different from test particle transport in a fluid, due to the strong
coupling between the electrostatic potential, which serves as the
stream function for the ExB fluid velocity, and the electron density
\cite{hasmim,wakhas,ssdw}.
The coupling is effected by the action of parallel currents and produces
a strong correlation between the potential and density.
Ultimately, if the electrons are adiabatic (Boltzmann relation vis-a-vis
the potential) then there is no net transport of electrons even though
there is turbulence, which could also be driven by the ion temperature
\cite{hortonestes,hamaguchi,dimits}.  
Since ExB transport is ambipolar, it follows that
there is also no net transport of ions (neglecting the small charge
density implied by a finite ExB vorticity).  Even in robustly
electromagnetic tokamak edge turbulence, where the free energy comes
from the extent to which the electrons are nonadiabatic, the adiabatic
response is always of qualitative importance, and there is a strong
correlation between the potential and the electron density
\cite{dalfloc,focus}.

What we will show herein is that in most reasonable circumstances the
test particle species, even when it is a tracer field with vanishing
influence upon the electrostatic potential, follows the same dynamics as
the electron density.  The process has a very similar appearance and
results ultimately from the same mechanism as ``dynamical alignment''
between a passively advected density and the vorticity in neutral fluid
turbulence \cite{gang}.  It is a manifestation of (1) the advective
nonlinearity being the most effective process in the equation for each
quantity, and (2) each of these quantities having a strong direct
cascade in its squared amplitude (playing the role of an entropy
contribution) despite the fact that the flow energy has an inverse
cascade \cite{sorgdw}.  The basic nonlinear dynamics is sufficiently
general that these cascade tendencies persist in the presence of strong,
dissipative forcing \cite{camargo}.  The result is that any initialised
imprint on the spatial morphology of either variable is quickly
transferred to small, dissipative scales, and the subsequent evolution,
being the same for both quantities, produces similar morphology
especially at scales within the inertial range of the turbulence.

The computations are done in three dimensional
flux tube geometry, with a fully self consistent three species gyrofluid
model \cite{beer} with two ion species plus electrons, generalised to
allow nonadiabatic, electromagnetic electron dynamics \cite{gyroem}, but
restricted to two moments per species in the ``GEM3'' model with clean
energetics and clear correspondence to fluid edge turbulence models
\cite{eps03}.
Not only the evolution of
the small scale eddies are followed, but also that of the zonal profiles
(flux surface averages, cf.\ Ref.\ \cite{beyer}).
This dynamics is followed to transport
equilibrium for cases driven by a source, and for at least one transport
decay time for cases with no sources.  The trace ion species is
initialised either the same as for the electrons (random bath
disturbances plus a profile), or simply with a narrowly localised
profile.  In some cases both ion species are given equal background
densities.  In each case, the initial differences are eliminated by the
evolution of the turbulence, and at late times the morphology of all
species densities are closely similar, with a high degree of cross
correlation between any two of them.

The main result is that even though the electrons are not transported
passively, the density of a trace species is transported the same way as
the electrons are.  A companion test is given by a simple two
dimensional drift wave model, following only the potential and electron
density and their dissipative coupling \cite{wakhas}, 
plus an additional equation for a third field which is the
density of a strict test particle species.  Here, the regime of the
electrons response is varied from deeply hydrodynamic to adiabatic
\cite{gang,sorgdw}.  In either case, the test
particle density closely follows the electron density, in all regimes,
regardless of whether the electron and test particle densities are
similarly or differently driven.
This should serve as a caution against prematurely concluding that
the density in a magnetised plasma are passively advected in the
case the electrons and a trace impurity species should be observed to
display the same transport properties.

Following sections document the GEM3 model as used, the results in
decaying and driven cases, respectively, and the results from the
dissipatively coupled, two dimensional model.  A discussion and summary
is given at the end.

\section{The Three Species GEM3 Model
\label{sec:gem3}}

For this study we use the GEM3 model, which allows a finite background
temperature for all species but does not follow the dynamics of the
temperatures or parallel heat fluxes \cite{eps03}.  It allows drift wave
and ideal and resistive ballooning dynamics according to the parameters
\begin{equation}
\bhat = {4\pi p_e\over B^2}\LP qR\over\Lpp\RP^2
  \qquad\qquad
  C=0.51{\nu_e\over c_s/\Lpp}{m_e\over M_i}\LP qR\over\Lpp\RP^2
\end{equation}
controlling the parallel electron response and
\begin{equation}
\delta = {\rs\over\Lpp} \qquad\qquad \wcv = 2{\Lpp\over R}
\end{equation}
controlling the perpendicular drift dynamics and the magnetic curvature
effects.  Both interchange and geodesic curvature are retained
\cite{gdcurv}.  
Normalisation is in terms of a profile scale $\Lpp$ and the sound speed
$c_s$, giving a natural drift frequency of $c_s/\Lpp$.  The parallel
dynamics is normalised in terms of $qR$, where the parallel connection
length is $2\pi q R$, giving the $qR/\Lpp$ scale ratios in the parameters.
The range of scales under consideration is everything between the drift
scale $\rho_s$ and the profile scale $\Lpp$.
The collisional drift wave regime is roughly $C>1$,
becoming electromagnetic if $\bhat>1$.
Ideal and resistive ballooning enter if 
$\bhat\wcv>1$ or $C\wcv>1$, respectively.  The first two conditions are
usually satisfied but the latter two are not.  Hence, the principal
physical process is drift Alfv\'en turbulence \cite{dalfloc,focus}.
Geodesic curvature is always
important as its role is to regulate the attendant zonal ExB flows
\cite{gdcurv}.

The profiles as well as the fluctuations in the dependent variables are
followed, so that transport as well as turbulence can be computed, even
though the equations are still homogeneous in the normalised parameters.
Each species is
characterised by a background charge density, temperature/charge ratio,
and mass/charge ratio, given by the normalised parameters
\begin{equation}
a_z = {n_z Z/n_e} \qquad\qquad
\tau_z = {T_z/ZT_e} \qquad\qquad
\mu_z = {M_z/ZM_D}
\end{equation}
in terms of a reference electron temperature $T_e$ and deuterium mass
$M_D$.  The scale ratio for the parallel dynamics comes in as
$\eps_z = \mu_z(qR/\Lpp)^2$.
For electrons, $a_e = \tau_e = -1$ and $\mu_e = -(m_e/M_D)$.  
For a single component deuterium plasma we would have
$a_i = \mu_i = 1$ and $\tau_i$ the nominal ion/electron temperature
ratio $T_i/T_e$.  With two ion species (the second labelled 't')
we specify the three parameters
independently for each species but then restrict to a neutral plasma by
taking $a_i = 1-a_t$.  For simplicity we will assume a deuterium/tritium
plasma with equal temperatures for all species.

The GEM3 model is given by moment equations for the density and parallel
velocity of each species,
\begin{equation}
\dztt{\nzfl} = -B\dpl{\uzfl\over B} 
	+ \kappacv\LP\phigfl+\tau_z\nzfl\RP
\end{equation}
\begin{equation}
\bhat\ptt{\Afl}+\eps_z\dztt{\uzfl} + C\jfl 
	= -\dpl\LP\phigfl+\tau_z\nzfl\RP 
	+ \kappacv\LP\eps_z\tau_z\uzfl\RP
\end{equation}
including the effects of the toroidal magnetic drifts also in the
parallel velocity moment equation
\cite{beer,gem}.
The ExB advective and parallel derivatives are given by
\begin{equation}
\dztt{} = \ptt{} + [\phigfl,] \qquad\qquad
  \dpl = {1\over B}\pss{} - [\bhat\Afl,]
\end{equation}
in terms of a Poisson bracket defined by
\begin{equation}
[f,g] = \delta\LP\pxx{f}\pyy{g}-\pyy{f}\pxx{g}\RP
\end{equation}
noting the species-dependent nature of $\phigfl$.
The curvature operator and perpendicular Laplacian are given by
\begin{equation}
\kappacv = \delta\wcv\LB\LP\cos s+g^{xy}_k\sin s\RP\pyyk{}+\sin s\pxx{}\RB 
\qquad
  \ddpp = \delta^2\LB\LP\pxx{}+g^{xy}_k\pyyk{}\RP^2+\ppyyk{}\RB
\label{eqkappacvdef}
\end{equation}
where $g^{xy}_k=\shat[s-s_k]$ 
is the off diagonal metric element and $\shat$ is the
shear, and 
$y_k$ denotes the member of the family of field aligned coordinate
systems which is orthogonal at $s=s_k$.
The gyroaverage operator is defined by
\begin{equation}
\phigfl = \Gamma_0^{1/2}(\kkpp\rho_z^2)\phifl \qquad\qquad
\rho_z^2 = {\mu_z\tau_z\over B^2}
\end{equation}
with $\rho_z$ the species gyroradius in terms of the nominal $\rho_s$
and with $\kkpp$ the Fourier transform of $-\ddpp$ 
(note $\kkpp$ contains the factor $\delta^2$).
The electrostatic potential is determined by the
gyrofluid analog \cite{dorland}
of the gyrokinetic Poisson equation \cite{wwlee},
\begin{equation}
\sum_z a_z\LB\Gamma_0^{1/2}\nzfl
	+ {\Gamma_0-1\over\tau_z}\phifl\RB = 0
\label{eqpol}
\end{equation}
with $\Gamma_0$ always taking the argument $(\kkpp\rho_z^2)$.
The magnetic potential is determined by the gyrofluid Ampere's law,
\begin{equation}
-\ddpp\Afl = \Jfl \qquad\qquad \Jfl = \sum_z a_z\uzfl
\end{equation}
hence by the various velocity moments making up the parallel current.

All of this is in local flux tube geometry defined separately with
respect to each drift
plane ($xy_k$, defined globally, orthogonal at $s=s_k$), 
each on its own globally field aligned coordinate system,
so that the linear parallel derivative ($\pps{}$)
incurs shifts in the $y$-direction to reflect the magnetic shear.  The
simplest version of the geometry is used, in which the metric of the
coordinate system referenced to each drift plane is unit diagonal on
that plane, and the magnetic field strength is $B=1$, giving the above
Laplacian and curvature operators.  This is the ``shifted metric''
treatment; for further detail see Ref.\ \cite{shifted}.  Elsewhere, the
subscript on $y_k$ is omitted (the form of the Poisson bracket is
unaffected). 

The equations are solved on a domain $L_x\times L_y\times L_s$, with
$L_s=2\pi$ due to the global consistency constraint.
The boundary conditions on the dependent
variables in the drift plane are Dirichlet (vanishing value)
at $x=L_x/2$ and Neumann (vanishing derivative) at $x=-L_x/2$, and
periodic on $-L_y/2<y<L_y/2$, with the $x$-direction
implemented in terms of the quarter-wave FFT so that the correct value
of $\Gamma_0$ is obtained for each wavenumber.  The boundary conditions
on $s$ are defined by global consistency \cite{fluxtube}, on the
interval $-\pi<s<\pi$, implemented using the shifted metric
treatment \cite{shifted}.

The numerical scheme is one which has been used previously in three
dimensional drift wave computation \cite{tyr}.  Poisson bracket
structures in $xy$ are evaluated with a discretisation which preserves
their properties \cite{arakawa}.  The linear $\pps{}$ and $\kappacv$
terms are evaluated using centred differences.  The timestep is a third
order scheme using both the dependent variables and their time
derivatives to build the new dependent variables at the next time step
\cite{karniadakis}.  In contrast to most other methods, this scheme is
stable to waves.  Nevertheless, the direct thermal free energy cascade
\cite{gang,sorgdw} creates the need for numerical dissipation at small
scale \cite{focus}.  This is effected by a diffusion $-\vpl\ppps{}$ and
a hyperdiffusion $\vpp\dpp^4$ added to each ExB derivative (applying the 
dissipation to the moment variables but not directly to the
fields).  The spatial grid node count is $64\times256\times16$ in $xys$,
on a domain given by $L_y=4L_x=80\pi\rs=4\Lpp$, or simply $4$ in
normalised units, and $L_s=2\pi$.  The minimum value of
$k_y\rho_s$ is therefore $0.025$.
The
timestep is $\tau=0.05$, and runs are taken to approximately $t=2000$.
The artificial dissipation coefficients are $\vpl=0.003$ and
$\vpp=0.002$.

\section{GEM3 Results for Decaying Cases
\label{sec:decaying}}

The first set of results under consideration is the decaying scenario:
the domain is loaded with a density profile and density fluctuations, 
and the system is allowed to relax.  Profile maintenance is restricted
to a sink term on the right hand side of
each density equation,
\begin{equation}
\ptt{\nzfl} = \cdots - \oint{dy\over L_y}\,\nzfl\,p(x)
\qquad\qquad
p(x)=0.3\exp\LB -{(x-L_x/2)^2\over (0.1\, L_x)^2}\RB
\end{equation}
where 
the integration denotes an average over $y$.  Net transport in the
direction $\grad x$ into this sink
then causes slow decay of the overall density
profile (zonal average of $\nefl$).

The initial profiles
are given by
\begin{equation}
n_{i0}= \half\LB 1-\sin{\pi x\over L_x}\RB \qquad\qquad
n_{t0}=\exp\LB -{x^2\over (0.1\, L_x)^2}\RB
\end{equation}
with
\begin{equation}
n_{e0} = (1-a_t) n_{i0} + a_t n_{t0}
\end{equation}
with subscripts $\{e,i,t\}$ referring to
electrons, main ions, and trace ions, respectively, with $a_t$ the
background charge density parameter of the trace ions.
The electrons and main ions start with a random bath of fluctuations at
nonlinear amplitude, $\nfl$, normalised such that the RMS level is
$3.0\times\delta$, described elsewhere \cite{ssdw,shifted}, with
\begin{equation}
\nefl = (1-a_t) \nfl \qquad\qquad \nifl = \nfl 
\qquad\qquad \ntfl = 0
\end{equation}
The trace ions
start solely with their profile.  
The initial electrostatic potential is
then given by Eq.\ (\ref{eqpol}).

The initial evolution
is the excitation of ``Pfirsch-Schl\"uter'' profile modes, especially
the global Alfv\'en oscillation.
This is
caused by the fact that the initial
equilibrium is only one-dimensional while the actual one is
two-dimensional, including Pfirsch-Schl\"uter currents and flows.
It involves the ``sideband
modes'' given by $\avg{\Jfl\cos s}$ and $\avg{\phifl\sin s}$, where the
angle brackets denote the zonal average (over $y$ and $s$).  These are
excited by the action of the magnetic compression $\kappacv$ upon the
pressure profile $\avg{p}$, where $p=\sum_z a_z\tau_z\nzfl$, due to the
$\sin s$ term in $\kappacv$ acting upon the $k_y=0$ component; 
in other words, the geodesic curvature.
The decay of the global
Alfv\'en oscillation proceeds through resistive dissipation, 
$C\avg{\Jfl\cos s}$.  
When these sideband modes reach dissipative
equilibrium, the Pfirsch-Schl\"uter current is established.  Over about
the same time interval, $0<t<200$, the basic nonlinear drift wave mode
structure is set up \cite{dalfloc,focus,eps03}.  
Parallel sound wave dynamics, the
geodesic acoustic sideband dynamics, and the zonal flows all reach
statistical equilibrium by about $t=1500$ \cite{gdcurv}.  For typical
edge parameters the profile decay (i.e., transport) time scale is
comparable to this, so we are dealing with a slowly transient case.
However, it does provide a useful control against the effects of profile
maintenance (driving of the densities using a source) upon the
turbulence.

The standard case, reflecting typical tokamak edge conditions, is given
by the parameters $\delta=0.0159$, $\bhat=1$, $C=2.5$,
$\tau_i=1$, $\wcv=0.05$, and $(qR/\Lpp)^2=18350$.  This very roughly
corresponds to a physical parameter set of $n_e=3\times 10^{13}\invcc$,
$T_e=100\eV$, $B=2.5\tesla$, $q=3.3$, $R=160\cm$, and $\Lpp=4\cm$.
As noted above, we also use deuterium as the main ion species and
tritium as the trace species, with $\tau_t=1$ and $\mu_t=1.5$ and
$M_D/m_e=3670$, and either $a_t=0$ (strict test
particles) or $a_t=0.1$ (minority species).  

\begin{figure}[H]
\includegraphics[width=\the\hsize]{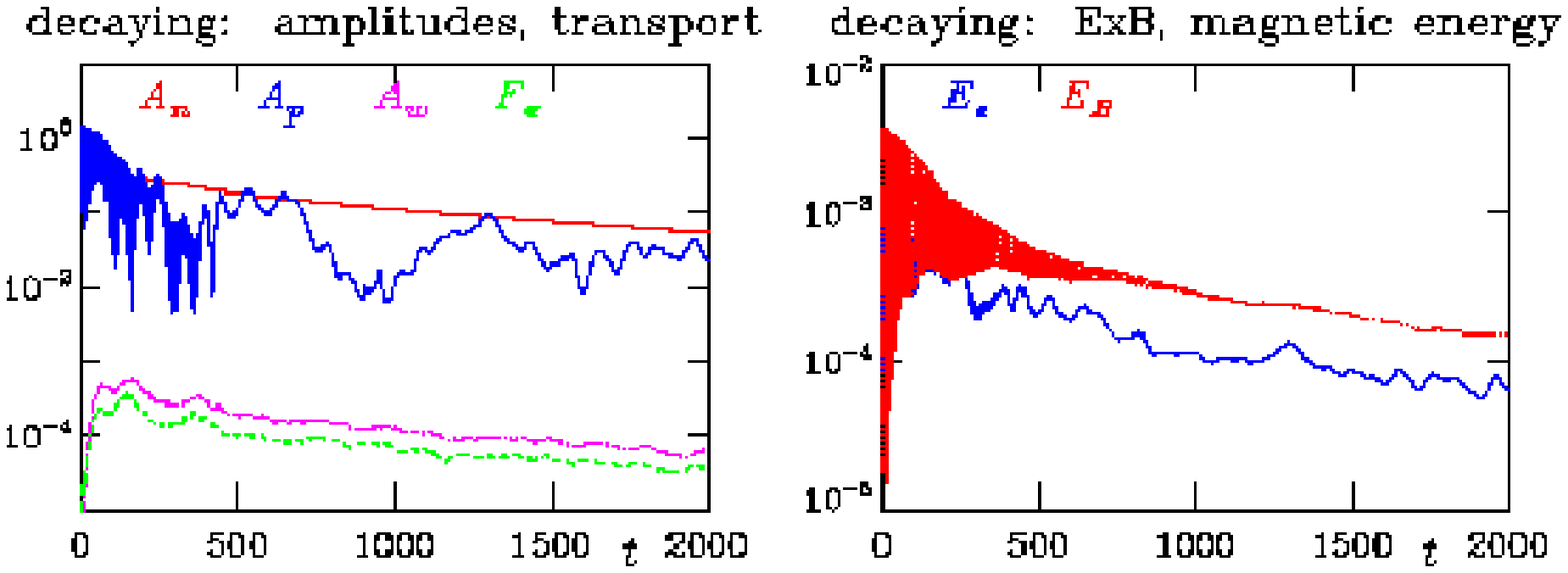}
\caption{
Time traces of free energy amplitudes and transport, showing initial
saturation
of the turbulence after $t=500$ and overall decay on a scale of about
$1500$, all in units of $\Lpp/c_s$.
}
\label{figdecaya}
\end{figure}

Selected time traces of the case with $a_t=0$ are shown in Fig.\
\ref{figdecaya}.  In the left hand frame, 
the signal $A_n$ is the grid node average of 
$\sum_z a_z\tau_z\nzfl^2/2$, giving the total thermal free energy,
and $A_p$ is $\phifl^2/2$, tracking mostly the zonal flows,
and $A_w$ is $\vorfl^2/2$, tracking mostly the activity of the turbulence,
and $F_e$ is $\nefl\vex$, the transport, with $\vex=-\delta(\ppy{\phifl})$.
In the right hand frame, the signal $E_e$ is $\sum_z a_z\phigfl\nzfl/2$,
the ExB energy, and $E_B$ is $\sum_z a_z\Afl\uzfl/2$, the magnetic energy.
Here, $\vorfl=-\sum_z a_z\nzfl$ defines
the generalised vorticity (at zero FLR it is just the ExB vorticity).
The signature of $E_e$ and $E_B$ is that of the
global Alfv\'en oscillation.
The
equilibrium is well established after about $t=250$.
The other traces show that the overall statistical equilibration time 
of the turbulence and zonal flows is comparable to the
transport decay time, and that the vorticity and transport track each
other closely.
By $t=1500$ the total amount of electrons, $\nefl$ integrated over the
volume, has decayed by about half.

\begin{figure}[H]
\includegraphics[width=\the\hsize]{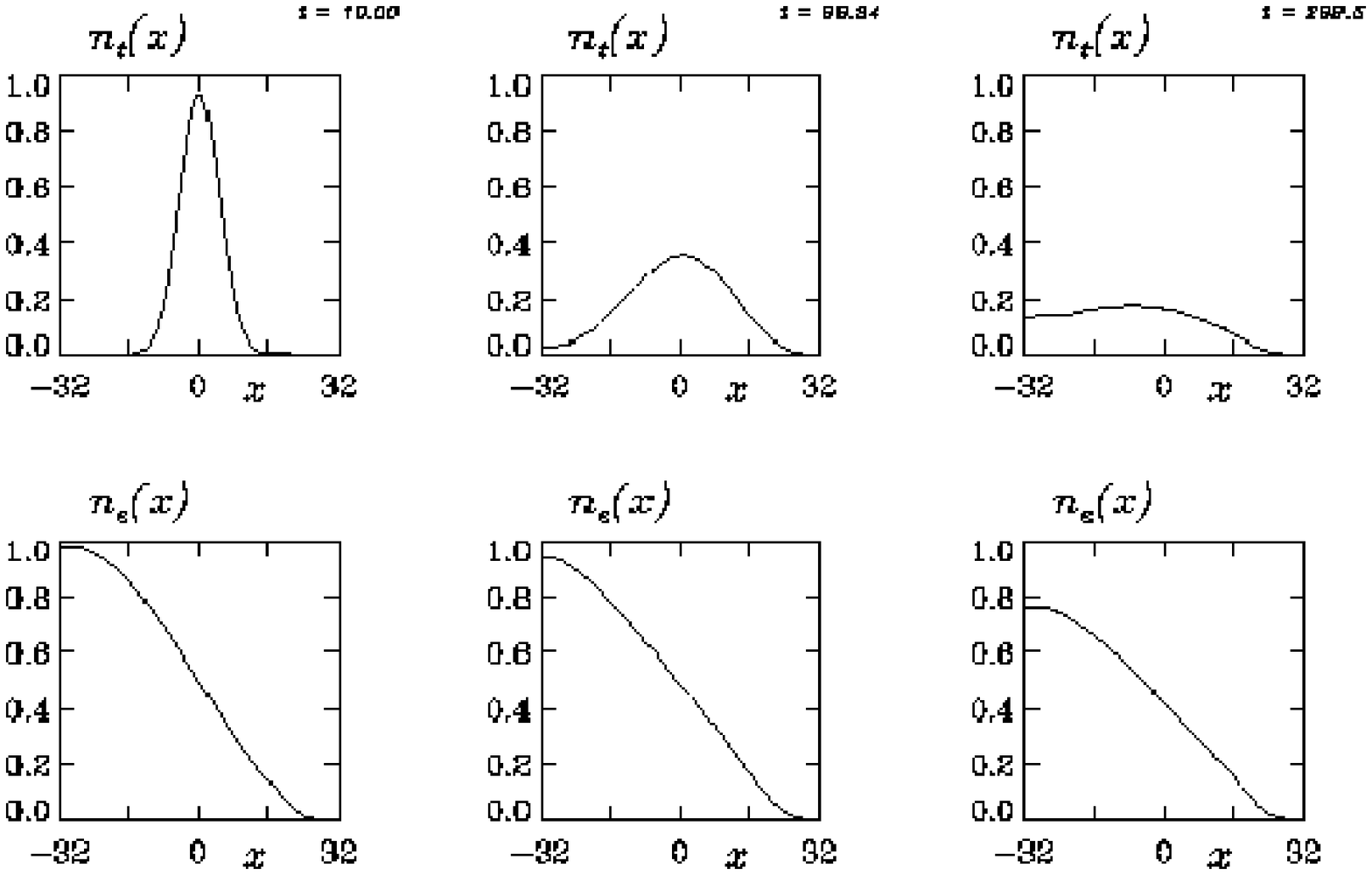}
\caption{
Profiles of the electron ($\nefl$) and trace ion ($\ntfl$) densities at
early times $t=10$ and $100$ and $300$
in the evolution.  The initial value is unity in these
units, and $x$ is in units of $\rs$.  The trace ion profile is
deconstructed by the turbulence and fills in.  
}
\label{figdecaypa}
\end{figure}

\begin{figure}[H]
\includegraphics[width=\the\hsize]{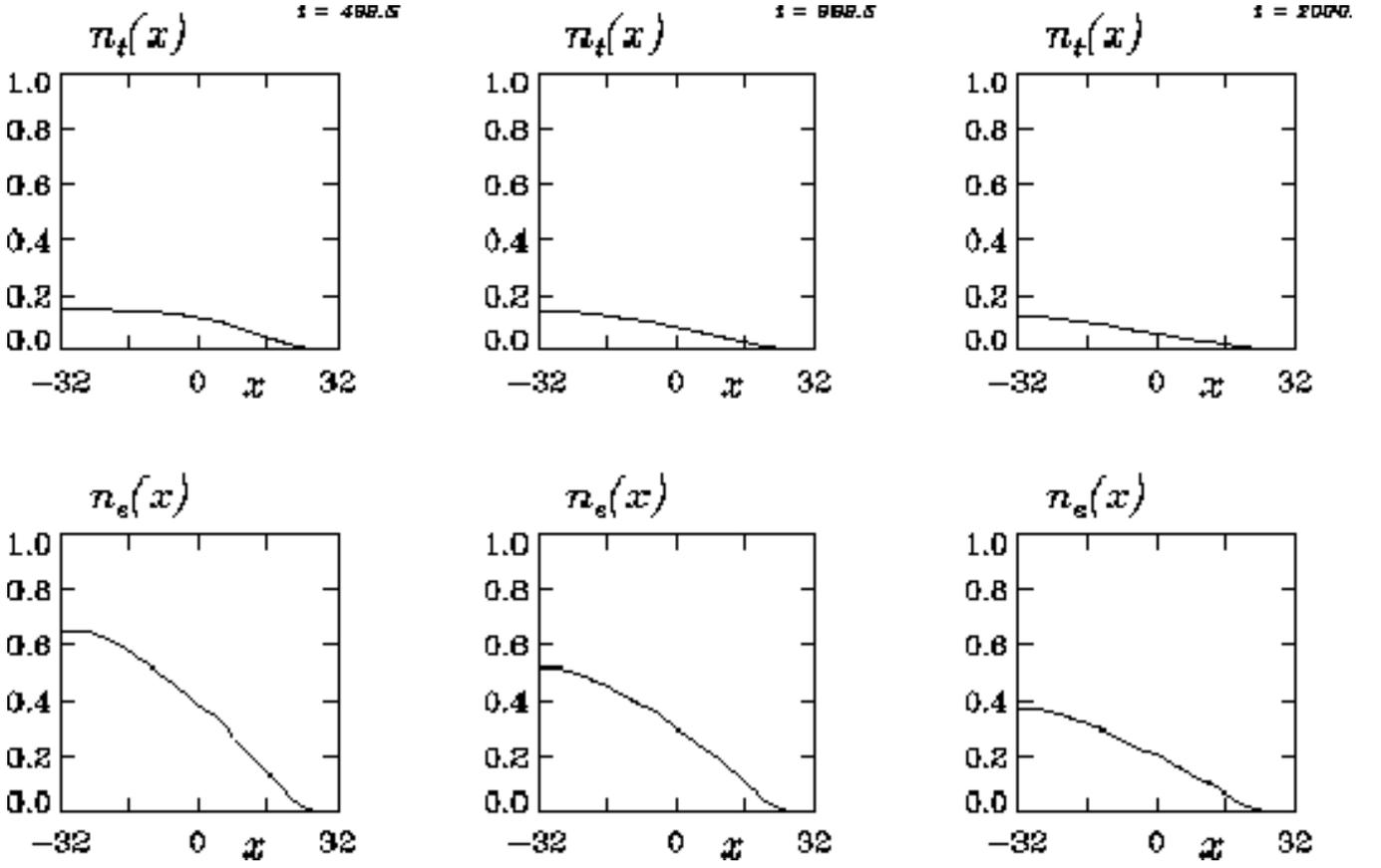}
\caption{
Profiles of the electron ($\nefl$) and trace ion ($\ntfl$) densities at
medium to late times $t=500$ and $1000$ and $2000$
in the evolution.  The initial value is unity in
these 
units, and $x$ is in units of $\rs$.  When the value of
the trace ion profile at the left boundary becomes the maximum, both
profiles decay together.
}
\label{figdecaypb}
\end{figure}

The evolution of the decaying profiles (zonal averages, over all $y$ and
$s$ for particular $x$) of the electron and trace densities is shown in
Figs.\ \ref{figdecaypa} and \ref{figdecaypb}.  In these units, the
maximal initial value is unity, and the first frame at $t=10$ shows that
the turbulence is already deconstructing the initial profile of $\ntfl$.
The trace species distribution spreads until the value at $x=-L_x/2$
($-32$ in units of $\rs$) rises from zero at $t=100$.  Thereafter
the trace profile fills in as the electron profile decays.  When the
value at $x=0$ reaches that at $x=-L_x/2$, the $\ntfl$ profile itself
begins to decay.  At late times, the profiles of $\nefl$ and $\ntfl$ are
similar.  As the driven cases discussed in the next Section will show,
this behaviour is not indicative of a particle pinch in either species.

\begin{figure}[H]
\includegraphics{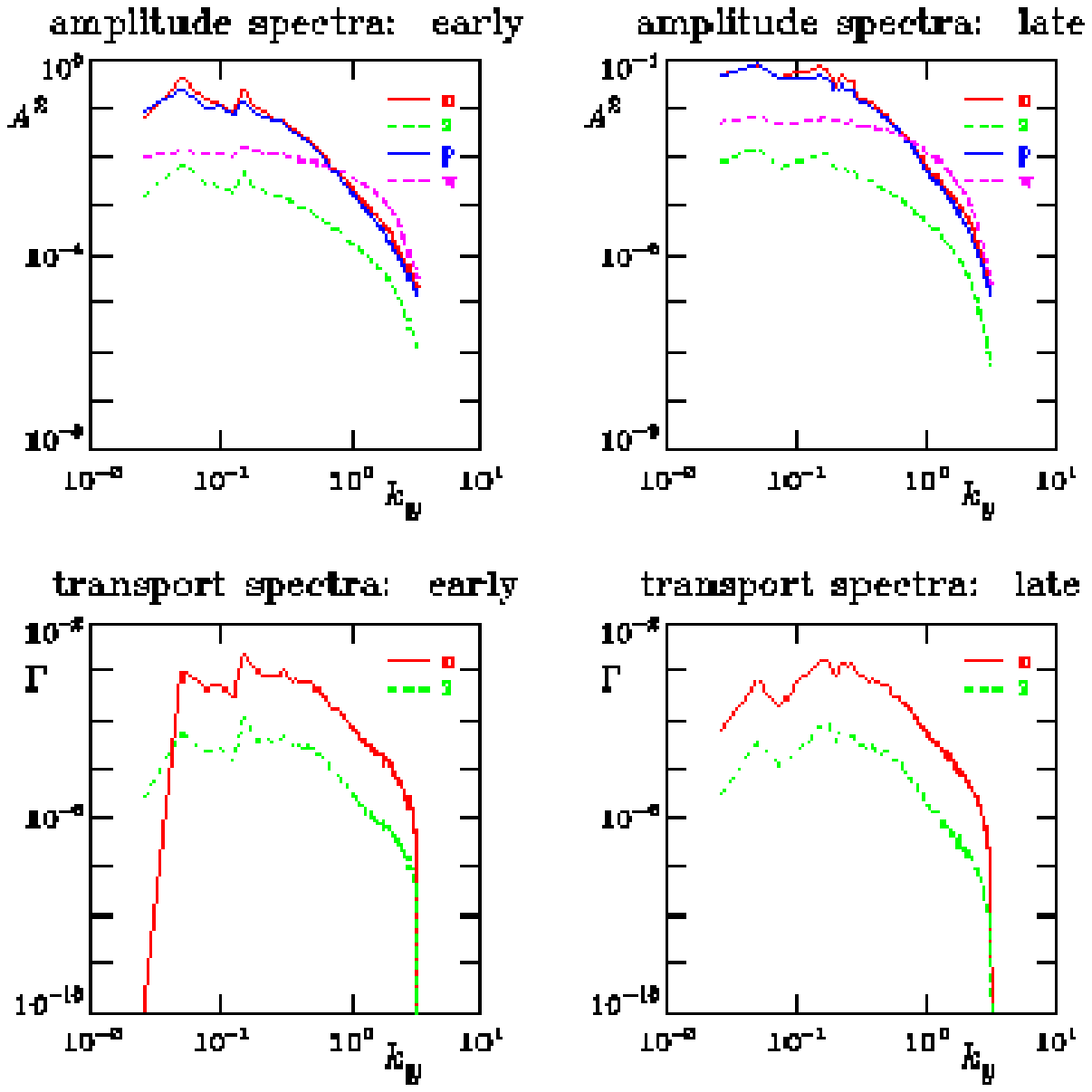}
\caption{
Spectra of selected quantities at early ($t=500$) and late ($t=1500$)
stages of the turbulence, with $k_y$ in units of $\rs^{-1}$.
Squared amplitudes of $\phifl$ and $\nefl$ and $\ntfl$
and the vorticity $\vorfl$ are labelled by `p' and `n' and `z' and `w'
respectively.  
The tendency of $\phifl$ and $\nefl$ to follow each other
with $\vorfl$ much flatter is a basic
signature of drift wave turbulence.  
The electron (`n') and trace ion (`z')
transport fluxes also follow each other.
}
\label{figdecayb}
\end{figure}

The most prominent amplitude/energy spectra (squared amplitudes, Fourier
decomposed $y\to k_y$, averaged over $0<x<L_x/2$ and all $s$) are shown
in Fig.\ \ref{figdecayb}, averaged over the intervals 
$500<t<600$ and $1500<t<1600$, respectively representing the early
and late stages of the turbulence.
These are reflective of typical drift wave mode structure as we
expect in this regime; note especially that $\nefl^2$
as a function of $k_y$ (labelled $n$) follows
$\phifl^2$ (labelled $p$) throughout the spectrum, as the average value
of the parallel wavenumber $\kpl$ self adjusts to whatever is necessary
for the parallel currents to balance the nonlinear excitation rates of
the ExB vorticity.  The vorticity spectrum itself,
$\vorfl^2$ (labelled $w$), is much flatter, extending all the way to
$k_y=1$ in units of $\rho_s^{-1}$.
The transport spectrum, $\nefl\vex$, also shown for both early and late
stages, peaks at relatively low $k_y$ but is broadband.
We find similarly standard drift wave mode structure
results for the parallel envelopes and transport and energy transfer
spectra, as presented in Refs.\ \cite{focus,eps03}, not shown here.

\begin{figure}[H]
\includegraphics[width=\the\hsize]{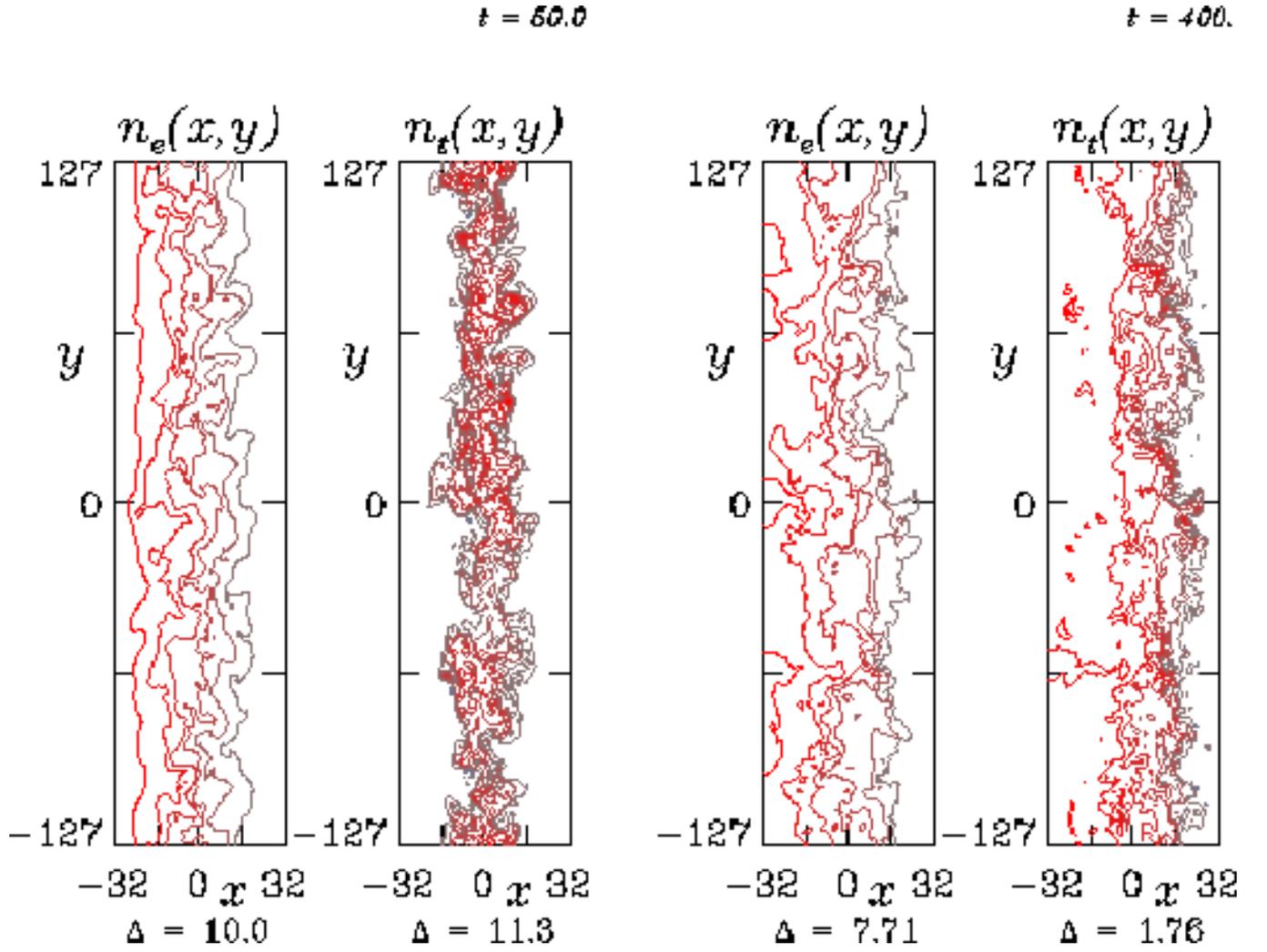}
\caption{
Spatial morphology of the electron ($\nefl$) and trace ion ($\ntfl$)
densities at $t=50$ and $400$, during saturation.  
The contour interval is shown in units of $\delta=0.0159$,
and $x$ and $y$ are in units of $\rho_s$ in each case.
Starting very different, their spatial forms evolve towards each other.
}
\label{figdecayc}
\end{figure}

The spatial morphology at $t=50$ and $400$ is shown in
Fig.\ \ref{figdecayc}, as 
contour patterns in the $xy$-plane at $s=0$.
The form of $\ntfl$ and $\nefl$ start out very differently owing to the
initialisation.  The spatial morphology 
however becomes quickly similar after the
turbulence deconstructs the original profile of
$\ntfl$ (transfer free energy from $k_y=0$ to the entire
$k_y$-spectrum).  At $t=50$ the part of the contour distribution for
$x>0$ is already very similar, even though the values at $x=-L_x/2$ are
still close to zero.  
However, the spectra and even the spatial morphology
evolve toward each other on all scales, including the profile scale.  
After $t=400$ they remain very close.  The fact that the spectrum of
$\ntfl$ is slightly flatter on the smallest scales is reflected in the
somewhat noisier form of the morphology of $\ntfl$, seen most clearly in
the outermost contour.

\begin{figure}[H]
\includegraphics{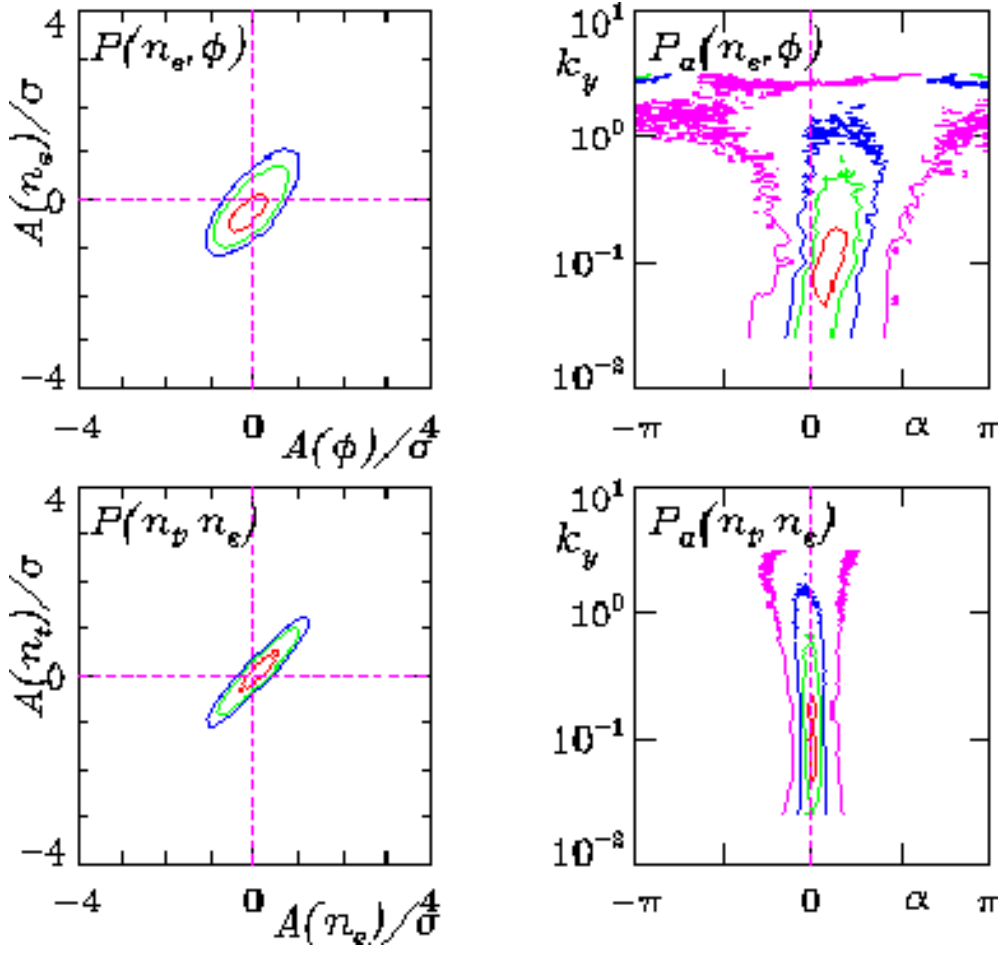}
\caption{
Cross correlation and phase shift distributions between 
the electron density ($\nefl$) and electrostatic potential ($\phifl$),
and between the trace ion density ($\ntfl$) and $\nefl$, 
averaged over the interval $500<t<1500$,
with $k_y$ in units of $\rho_s^{-1}$,
measured as described in Refs.\ \cite{dalfloc,focus}.
Note that $\phifl$ serves as the stream function for the turbulent
ExB eddies.
The first pair show the usual signature of drift wave mode structure
(close correlation and small phase shifts at all wavelengths), while the
second pair show the dynamical alignment between the trace ion and
electron densities.  This is the case with $a_t=0$; the one with
$a_t=0.1$ was quantitatively very similar in all respects, as detailed
in the text.
}
\label{figdecayd}
\end{figure}

The closeness of $\ntfl$ and $\nefl$ may be quantified by their cross
correlation and phase shift distributions.  These are more usually
employed to evaluate the relation between $\nefl$ and $\phifl$,
to distinguish between drift and MHD mode structure 
\cite{dalfloc,focus,eps03}).  An ``adiabatic'' relationship is
characterised by a close cross correlation and a narrow phase shift
distribution peaked near zero.  A ``hydrodynamic'' relationship, the one
expected of passive scalar advection in a neutral fluid, is
characterised by a very weak or zero cross correlation and a wide
phase shift distribution centred upon $\pi/2$, the value which gives
energetically maximum gradient drive.
The fields in question are sampled on the $xy$-plane at $s=0$, with the
$k_y=0$ mode stripped out, 
over the interval $500<t<1500$ with the turbulence saturated and
still well developed,
and for $5<x<27$ to keep clear of the outer boundary sink region and
remain in the gradient region for $\ntfl$.
These diagnostics are shown for both $\{\nefl,\phifl\}$ and 
$\{\ntfl,\nefl\}$ in Fig.\ \ref{figdecayd}.
The first pair show close correlation (overall: $0.721$) and low phase
shifts (positive, narrowly distributed, below $\pi/4$) indicative of
drift wave mode structure.  The second pair shows extremely tight
correlation (overall: $0.937$) and a narrow phase shift distribution
centred upon zero.

The expected form for a passive scalar quantity completely uncoupled to
the potential would be hydrodynamic, with random phase shifts in a case
with no gradient.  By contrast, although in the dynamical equations
there is no direct coupling between $\nefl$ and $\ntfl$, they are very
closely correlated and their phase shift distribution is narrow and
peaked at zero for every nonzero $k_y$ in the spectrum.  Taking the two
combinations from this figure together, we find that the electron and
trace ion density fluctuations track each other very closely, despite
the fact that the electrons themselves are not in a passive advection
relationship to the ExB eddies.  This is the phenomenon we refer to as
``dynamical alignment'' and the center of the phase shift distribution
(essentially zero) and the value of the overall cross correlation
(larger than $0.9$) give it a quantitative basis.

\begin{figure}[H]
\includegraphics{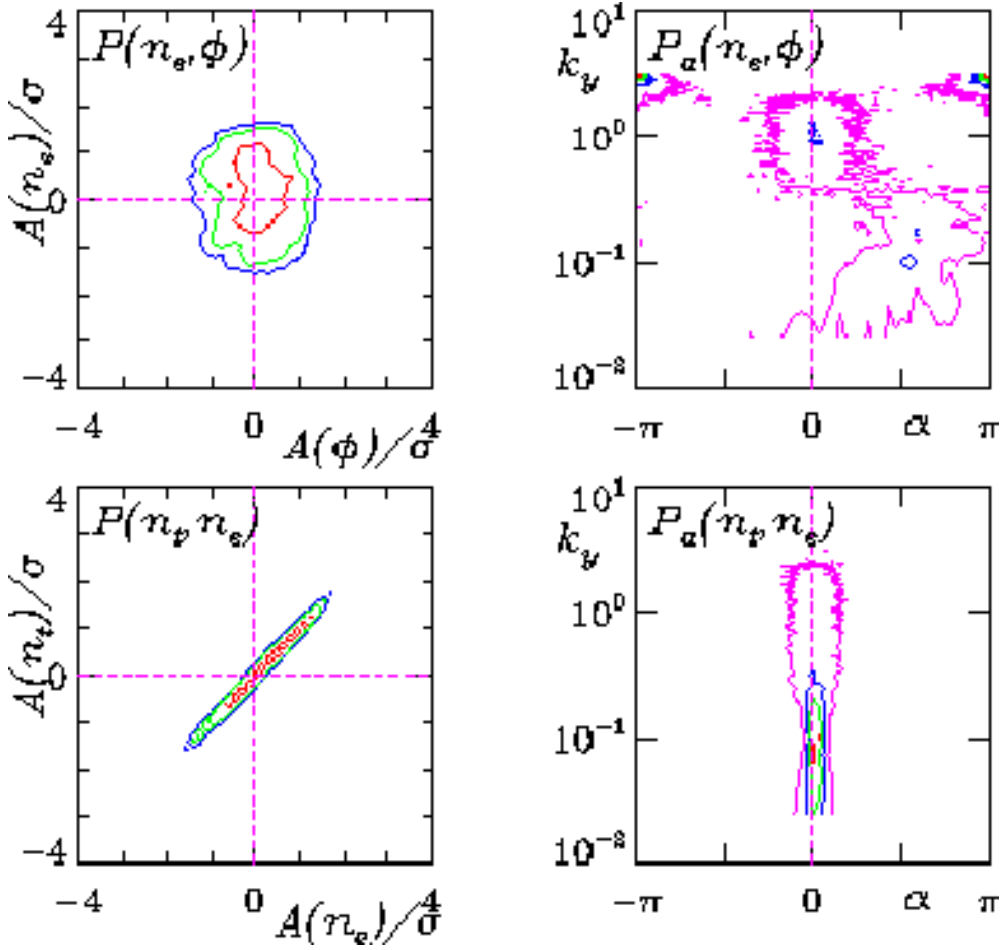}
\caption{
Cross correlation and phase shift distributions between 
the electron density ($\nefl$) and electrostatic potential ($\phifl$),
and between the trace ion density ($\ntfl$) and $\nefl$, 
for the case with $a_t=0.1$ and $\bhat=5$,
averaged over the interval $500<t<800$,
with $k_y$ in units of $\rho_s^{-1}$.
The first pair
now shows the high-beta ideal ballooning 
mode structure (very low correlation and phase shifts near $\pi/2$).
The second pair shows the 
dynamical alignment between the trace ions and the electrons.
Compare to Fig.\ \ref{figdecayd}.  The dynamical alignment persists
regardless of the actual character of the electron transport itself.
}
\label{figdecaye}
\end{figure}

Testing against the presence and absence of a back reaction of $\phifl$
to $\ntfl$, this basic case with $\bhat=1$ and $a_t=0$
was run again 
with $a_t=\{0.1,0.2,0.3,0.5\}$.  The same results as presented above for
$a_t=0$ were found, with the cross correlation values for $a_t=0.1$
given by $0.763$ for $\nefl$ and $\phifl$ and 
$0.947$ for $\ntfl$ and $\nefl$, respectively.
Cases with $\bhat=5$ with $a_t=0$ and $0.1$ were also run, 
looking for changes closer to the MHD
regime in which the mode structure should be more
hydrodynamic.  The basic mode structure shows the changes resulting from
the stronger role of the vortices at $k_y\rs=0.05$ to $0.1$ (convective
cell modes on the scale of $\Lpp$), which in the $k_y$-spectra sit
higher above the broadband drift wave turbulence and show much
larger phase shifts.  Nevertheless, the same closeness of $\ntfl$ to
$\nefl$ is found.  The cross correlation and phase shift information as
in the nominal case are shown in Fig.\ \ref{figdecaye} for the case with
$\bhat=5$ and $a_t=0.1$.  The fact that this case is in the MHD regime
is shown by the hydrodynamic relationship between $\nefl$ and $\phifl$;
the overall cross correlation value is $-0.0259$.  The persistence of
the dynamical alignment is shown by the relationship between $\ntfl$ and
$\nefl$, which gives zero phase shifts and a overall cross correlation
of $0.986$.

Regardless of the closeness of
the relationship between $\nefl$ and $\phifl$, which varies with
parameters (in this case $\bhat$ and $C$), the correlation between
$\nefl$ and $\ntfl$ is close to unity and the phase shifts are close to
zero.  Clearly, $\ntfl$ is following $\nefl$ regardless of the latter's
relationship to $\phifl$.  

What we now have to do is to determine the extent to which this result
might be a fortuitous consequence of how the model is set up.  To this
end, we compare to driven cases run to transport equilibrium in the next
Section, and then evaluate the phenomenon more fundamentally by studying
the simplest dissipative drift wave model in the adiabatic and
hydrodynamic limits, in the Section after that.

\section{GEM3 Results for Driven Cases
\label{sec:driven}}

We now turn to cases in which the profiles are maintained by sources
so that the
runs may be carried for arbitrary time.  The setup is exactly as for the
decaying cases, with the addition of the drive term
\begin{equation}
\ptt{\nzfl} = \cdots + S_z(x)
\label{eqsourcedef}
\end{equation}
for each density variable.  Typically we use $S_i=S_e$ so as not to
drive vorticity on the inner boundary.  
The basic profile source term has the
shape of a Gaussian with $1/e$-width $0.3L_x$ centred upon $x=-L_x/2$
and a time
constant chosen to anticipate the transport time scale.  Narrower main
source profiles (width $0.1L_x$)
with correspondingly larger coefficients (i.e., at
constant total source) were 
found to excessively excite artificial MHD modes.  
The trace ion source was centred upon $x=0$ to set up a gradient region
for $x>0$ and to look for a pinch (finite gradient with no interior
source) in the $x<0$ region, so $n_{t0}$ was used as a template.
The source terms were chosen as
\begin{equation}
S_e = S_i = s_e \exp\LB -{(x+L_x/2)^2\over (0.3\, L_x)^2}\RB
\qquad\qquad S_t = s_t n_{t0}
\end{equation}
for electrons, main ions, and trace ions, respectively.  The parameters
$s_e$ and $s_t$ are free; nominally they are both set to $10^{-3}$.

\begin{figure}[H]
\includegraphics[width=\the\hsize]{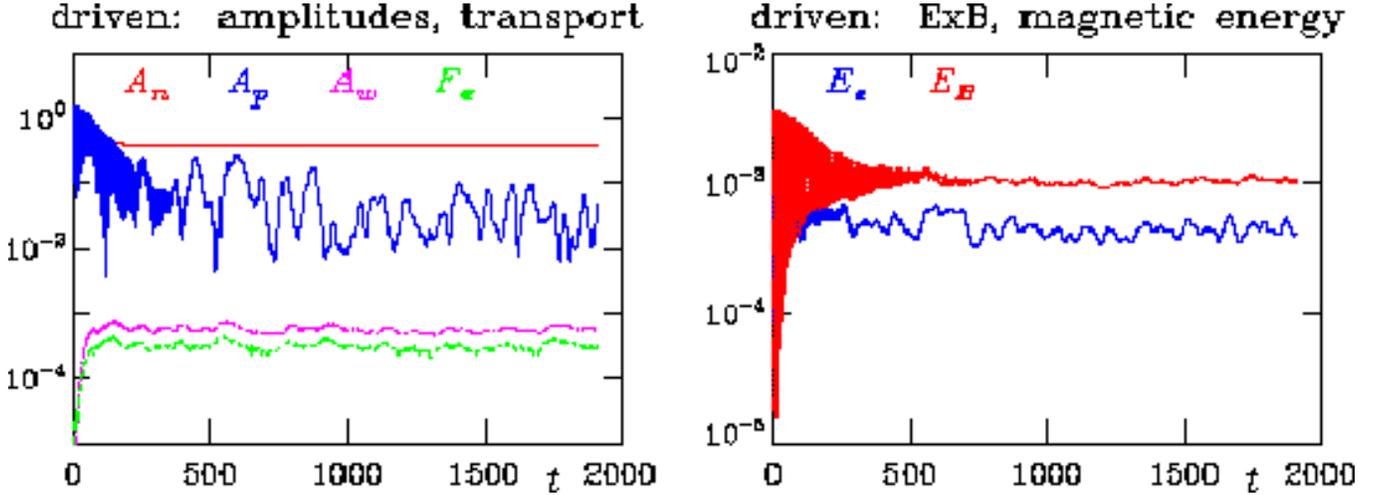}
\caption{
Time traces of free energy amplitudes and transport, showing initial
saturation
of the turbulence after $t=500$ and equilibration of the zonal flows and
transport after about $t=1000$, all in units of $\Lpp/c_s$.
}
\label{figdrivena}
\end{figure}

The standard case is the same as the one from the previous Section, with
$a_t=0$, so that the trace ions are proper test particles.  The same
time traces are shown in Fig.\ \ref{figdrivena}, which can be compared
to Fig.\ \ref{figdecaya}.  The amplitudes and transport, in the left
hand frame, show that the dynamics is well saturated after
$t=1000$, so that late time averaged diagnostics are taken over the
interval $1000<t<2000$.  The time traces of the ExB and magnetic
energies, labelled $E_e$ and $E_B$, respectively, are shown in the right
hand frame, indicating that the global Alfv\'en oscillation damps away
well before the saturated stage.  The short time scale visible in the
$A_p$ signal is indicative of the geodesic acoustic oscillation.
The long time scale oscillation from 
the decaying case is absent in this figure, showing that it is a
response to the overall profile decay in the other case.

\begin{figure}[H]
\includegraphics[width=\the\hsize]{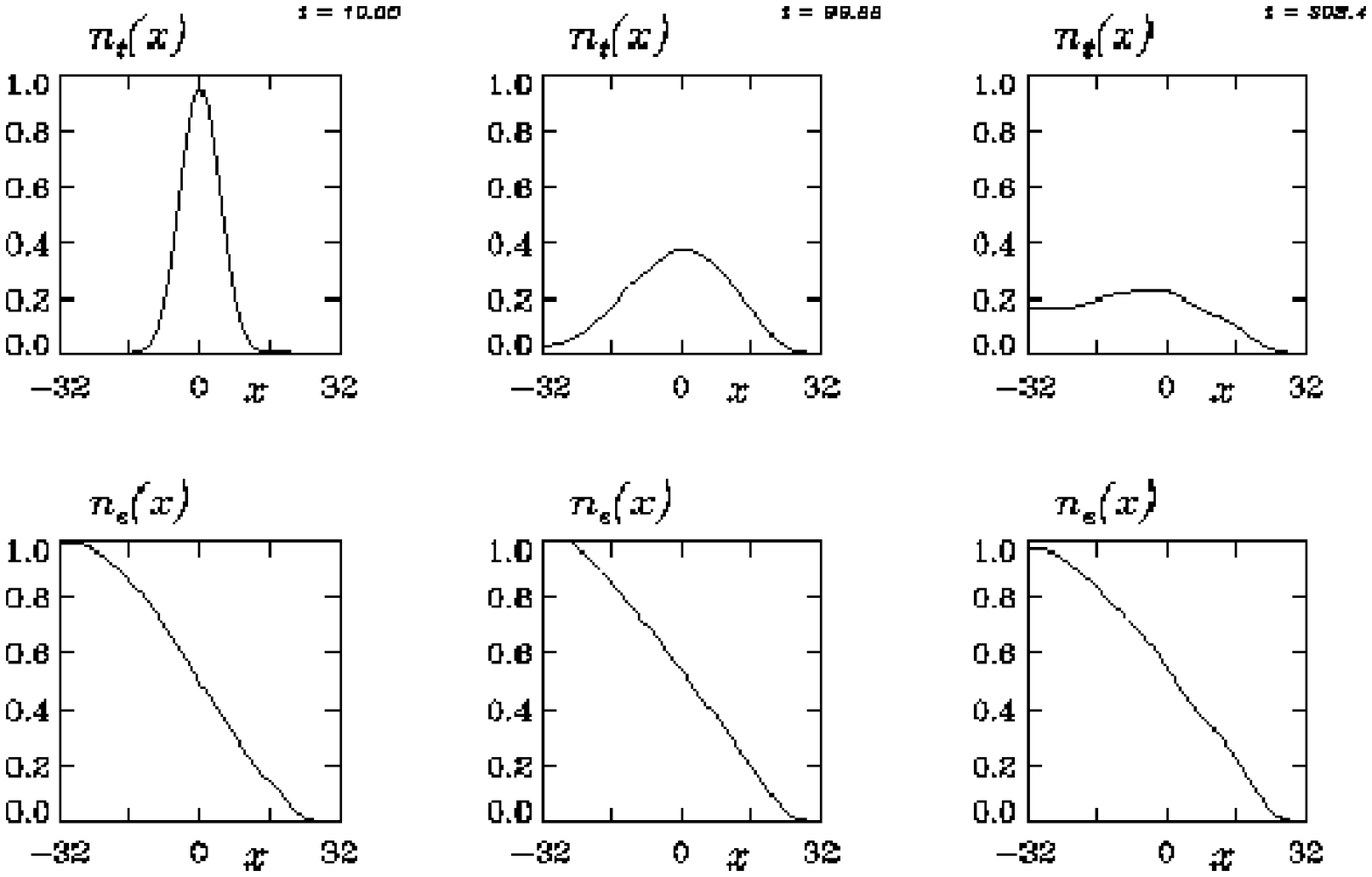}
\caption{
Profiles of the electron ($\nefl$) and trace ion ($\ntfl$) densities at
early times $t=10$ and $100$ and $300$
in the evolution.  The initial value is unity in these
units, and $x$ is in units of $\rs$.  The trace ion profile is
deconstructed by the turbulence and fills in.  
}
\label{figdrivenpa}
\end{figure}

\begin{figure}[H]
\includegraphics[width=\the\hsize]{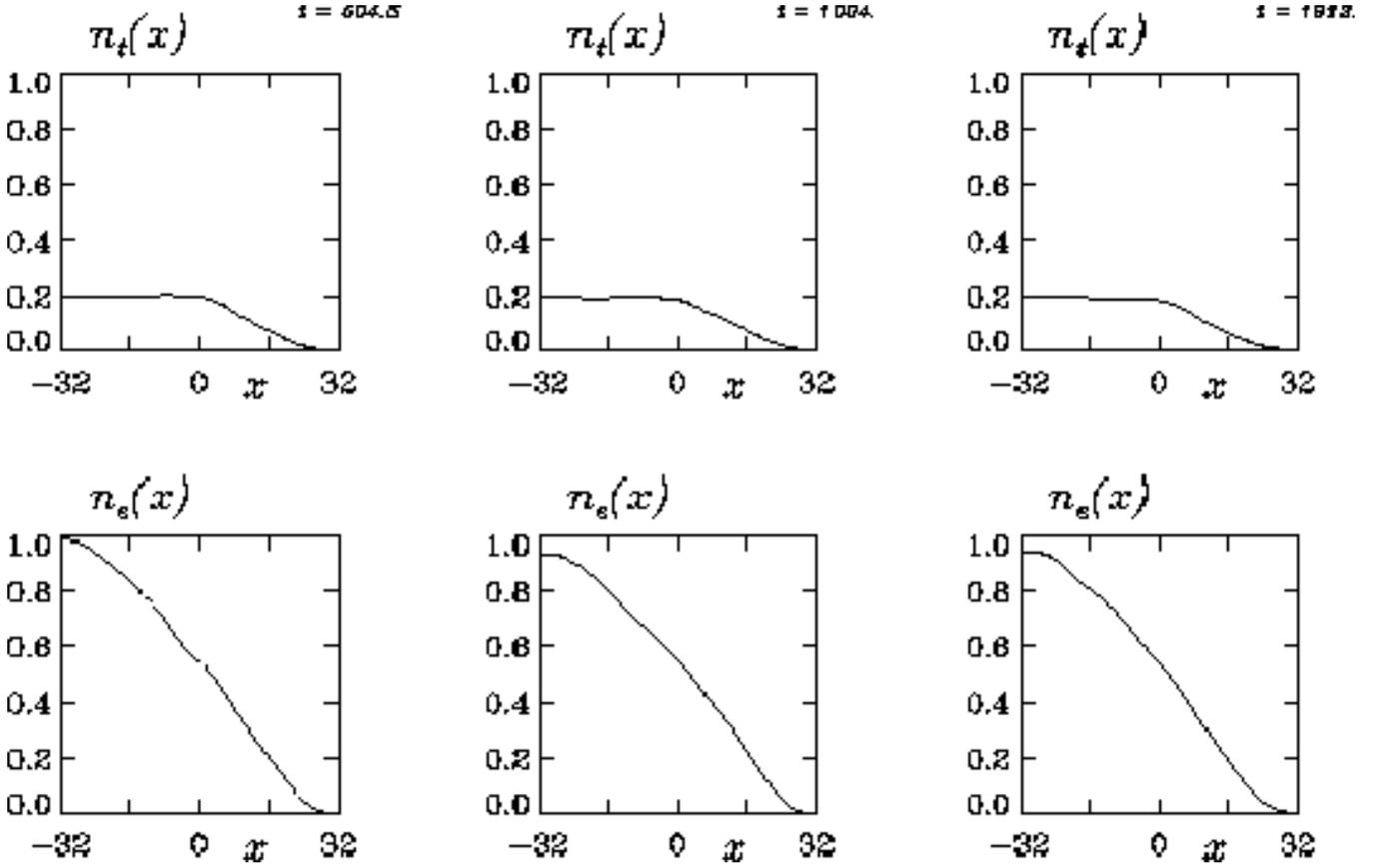}
\caption{
Profiles of the electron ($\nefl$) and trace ion ($\ntfl$) densities at
medium to late times $t=500$ and $1000$ and $2000$
in the evolution.  The initial value is unity in
these 
units, and $x$ is in units of $\rs$.  Transport equilibration occurs
after about $t=1000$.
}
\label{figdrivenpb}
\end{figure}

The evolution of the profiles of the electron and trace densities in
this 
driven case is shown in
Figs.\ \ref{figdrivenpa}\ and \ref{figdrivenpb}, 
measured the same way
as in Figs.\ \ref{figdecaypa}\ and \ref{figdecaypb}, 
for the decaying case.
In these units, the
maximal initial value is unity, and the first frame at $t=10$ shows that
the turbulence is already deconstructing the initial profile of
$\ntfl$.  Up until $t=100$ the evolution is the same as in the decaying
case.  
The trace species distribution spreads until the value at the left
boundary ($x=-32$ in units of $\rs$)
is as large as that at the source ($x=0$).
Transport equilibration, the temporal convergence of these
profiles, sets in after $t=1000$.
At late times, the profiles of $\nefl$ and $\ntfl$ are
similar outside the source regions.  The flatness of $\ntfl$ for $x<0$
is indicative of the absence of an impurity pinch.  For this we note
however that temperature disturbance and gradient dynamics is not
followed, so there remains the possibility of an impurity pinch driven
by the ion temperature gradient.
Nevertheless, the background temperatures of the three species are
equal, so this does show that the existence of warm ions does not by
itself lead to an impurity pinch nor does the trace and main ion
temperature lead to a ``neoclassical'' pinch, the underlying dynamics of
which would indeed be contained in this model (geodesic curvature acting
on the gradients through diamagnetic fluxes, and the interaction of this
with parallel flows in the axisymmetric part of the dynamics).

\begin{figure}[H]
\includegraphics{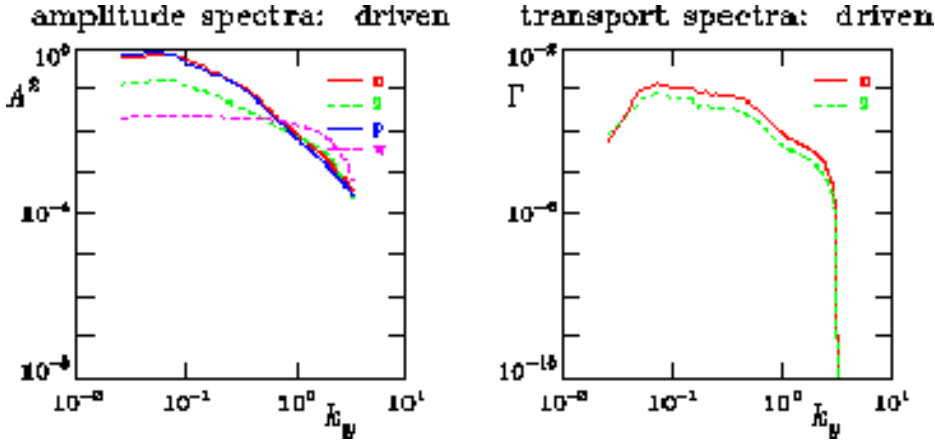}
\caption{
Spectra of amplitudes and transport in the saturated stage
$1000<t<2000$, with $k_y$ in units of $\rs^{-1}$. 
Squared amplitudes of $\phifl$ and $\nefl$ and $\ntfl$
and the vorticity $\vorfl$ are labelled by `p' and `n' and `z' and `w'
respectively.  
The tendency of $\phifl$ and $\nefl$ to follow each other
with $\vorfl$ much flatter is a basic
signature of drift wave turbulence.  
The electron (`n') and trace ion (`z')
transport fluxes also follow each other.
}
\label{figdrivenb}
\end{figure}

The spectra of the amplitudes and transport are shown in Fig.\
\ref{figdrivenb}, averaged over the saturated stage, given by
the interval $1000<t<2000$.  These show the amplitudes of $\nefl$ and
$\phifl$ peaking at large scales ($k_y<0.1$ in units of $\rs^{-1}$),
while the transport is broadband almost all the way to $k_y=1$.  The
vorticity is flat to $k_y=1$.  All of these are indicative of drift wave
mode structure.  The spectrum of $\ntfl$ follows $\nefl$.

\begin{figure}[H]
\includegraphics[width=\the\hsize]{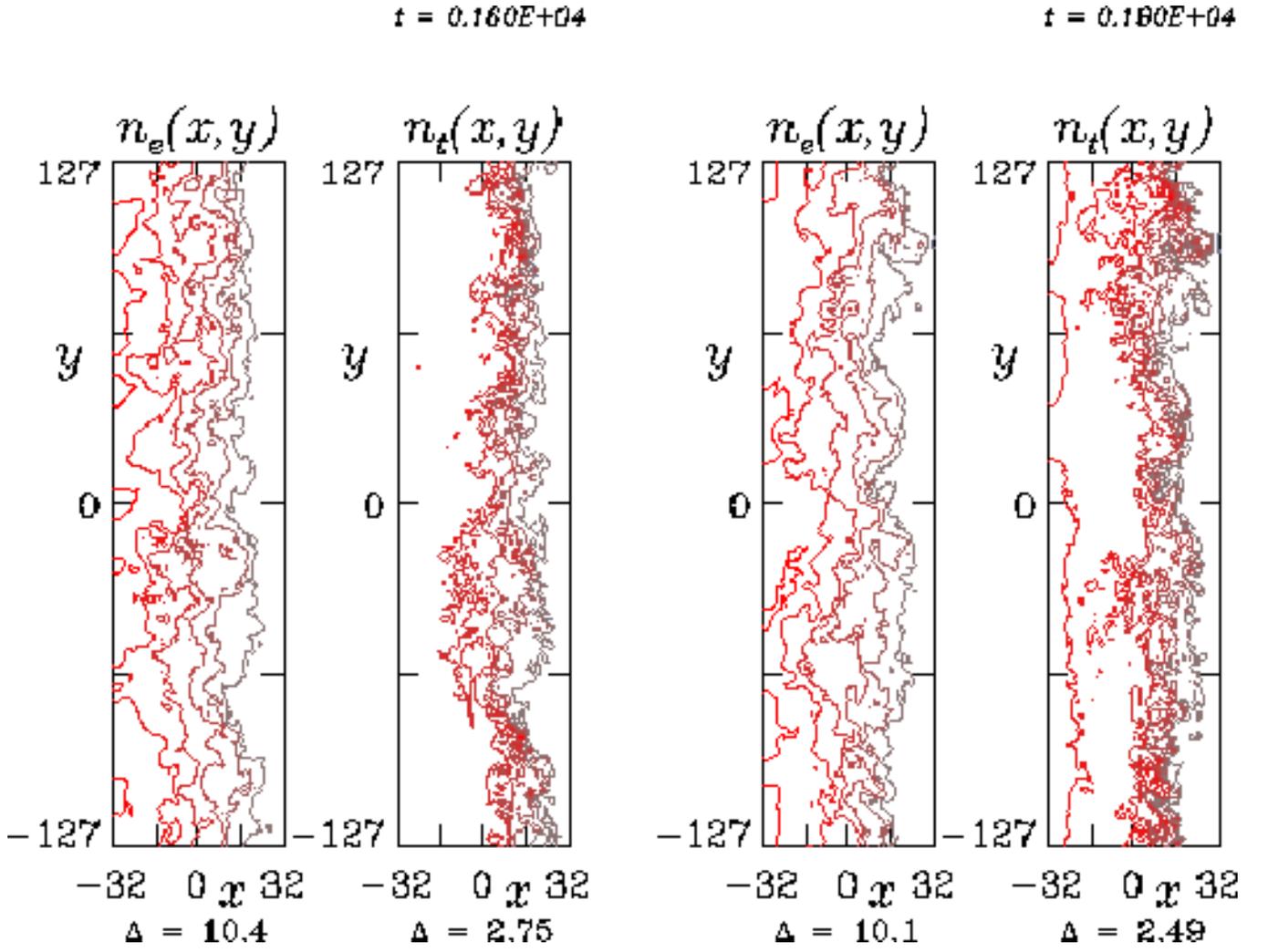}
\caption{
Spatial morphology of the electron ($\nefl$) and trace ion ($\ntfl$)
densities at two different times near the end of the run.
The contour interval is shown in units of $\delta=0.0159$,
and $x$ and $y$ are in units of $\rho_s$ in each case.
Starting very different, their spatial forms track each other closely in
saturation. 
}
\label{figdrivenc}
\end{figure}

The spatial morphology at $t=2000$ is shown in
Fig.\ \ref{figdrivenc}, as 
contour patterns in the $xy$-plane at $s=0$.
The form of $\nefl$ and $\ntfl$ are closely similar.
The fact that the spectrum of
$\ntfl$ is slightly flatter on the smallest scales is reflected in the
somewhat noisier form of the morphology of $\ntfl$, seen most clearly in
the outermost contour.  From time to time a small scale structure
appears in $\ntfl$ which is not reflected in $\nefl$, but the
occurrences are rare enough not to affect the statistical result.

\begin{figure}[H]
\includegraphics{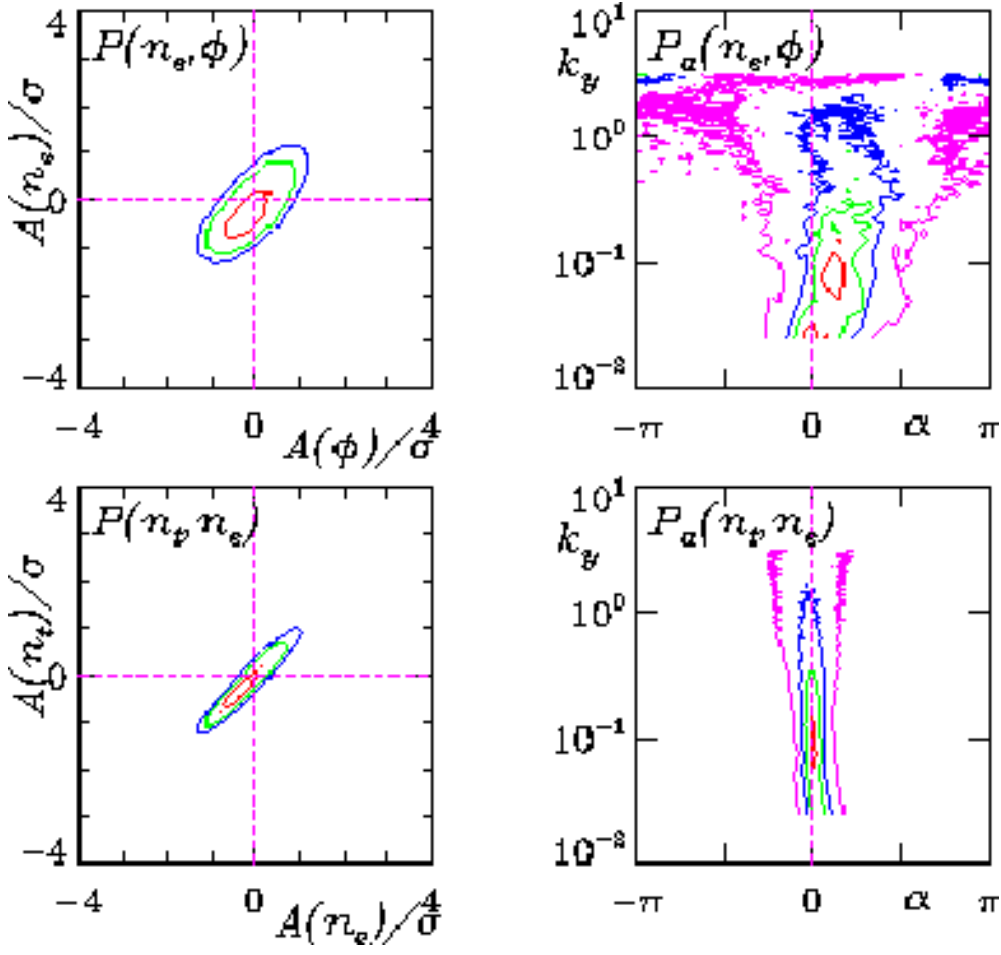}
\caption{
Cross correlation and phase shift distributions between 
the electron density ($\nefl$) and electrostatic potential ($\phifl$),
and between the trace ion density ($\ntfl$) and $\nefl$, 
for the driven case with $a_z=0.1$,
averaged over the saturated stage $1000<t<2000$,
with $k_y$ in units of $\rho_s^{-1}$,
measured as described in Refs.\ \cite{dalfloc,focus}.
Compare with Fig.\ \ref{figdecayd}.  Provided the sampling is done
exterior to the source region, the results are the same.
In particular, the drift wave mode structure signature remains
whether or not the turbulence is ``ambient'' or ``source driven''.
}
\label{figdrivend}
\end{figure}

The cross coherence and phase shifts between 
$\nefl$ and $\phifl$ and between
$\ntfl$ and $\nefl$ are
shown in Fig.\ \ref{figdrivend}, sampled over the saturated stage
the same way as in the decaying
case.  The first pair show that the signature of drift wave mode
structure (overall cross correlation: $0.674$)
is not affected by the proximity of the source.  
For realistic drive levels there is therefore
no qualitative difference between
``ambient'' and ``source driven'' turbulence, provided the physical
separation of scales is well reproduced in the computations (note that a
transport equilibration time of $1000$ with tokamak edge values of
$\Lpp/c_s$ is already very short).
The second pair show that the presence of the source just to the left of
the measured region does not affect the dynamical alignment result
(overall cross correlation: $0.920$), as can be judged by comparing with
Fig.\ \ref{figdecayd}.  But if this is measured over the whole radial
domain,
then the effect of not only the source but also the region interior to
it is found to be large.  So the conclusion of dynamical alignment
between these two disturbance fields does depend on the relative
strength of the nonlinear dynamics.

Having now determined that the three dimensional model recovers the same
basic result in source free gradient regions,
whether or not it is driven or allowed to decay, we now
turn to the simplest two dimensional dissipative drift wave model for
further comparison.

\section{Results in a Simple 2D Drift Wave Model
\label{sec:hwtp}}

We distill the central physics behind the above results by comparing
them to those from a simple two dimensional Hasegawa Wakatani drift wave
model \cite{wakhas}, augmented by a continuity equation for a test
species.  The equations are
\begin{equation}
\ptt{\vorfl} + \vedl\vorfl = D\LP\phifl-\nefl\RP
	- \kappacv\LP\nefl\RP + \nu_d\phifl
\end{equation}
\begin{equation}
\ptt{\nefl} + \vedl\LP\nefl+n_e\RP
	= D\LP\phifl-\nefl\RP
	+ \kappacv\LP\phifl-\nefl\RP
\end{equation}
\begin{equation}
\ptt{\ffl} + \vedl\LP\ffl+f\RP = 0
\end{equation}
for disturbances in the ExB vorticity, electron density, and test
particle density, respectively.  The domain is a single $xy$ drift
plane, doubly periodic.  Background gradient drive terms appear for both
the electrons ($n_e$) and test particles ($f$); their $x$-derivatives
yield the background gradients.
The ExB advective derivative and vorticity are given by
\begin{equation}
\vedl = \LP\pxx{\phifl}\pyy{}-\pyy{\phifl}\pxx{}\RP
\qquad\qquad
\vorfl=\ddpp\phifl
\end{equation}
The dissipative coupling parameter $D$ serves as a model for the
adiabatic response.  The hydrodynamic and adiabatic limits are $D\to 0$
and $D\to\infty$, respectively.  For $D=0$ the electrons and test
particles have the same equation.  For $D>1$ they are obviously very
different.  A simple interchange curvature model with
$\kappacv=\wcv\ppy{}$ and a long-wave damping coefficient $\nu_d$ are
included to study saturated interchange turbulence.
Note that this model uses the traditional ``gyro-Bohm'' local 
normalisation, with the factor of $\delta$ folded into the amplitudes.

For the purposes of this test series, the zonal flow question is avoided
by setting temporal changes to the $k_y=0$ component of $\vorfl$ to
zero.  The Colella \cite{colella}/Van Leer\cite{vanleer} MUSCL scheme is
applied to $d/dt$ as in previous versions of the 3D electromagnetic
gyrofluid model\cite{gyroem} (the linear drive terms are combined into
this, e.g., by defining $N=\nefl+n_e$).  The terms involving $D$ are
done via the implicit scheme of previous 2D drift wave models
\cite{jcp}.  In this, $D$ is applied only to the $k_y\ne 0$ components.
The background profiles are chosen differently, with $\ppx{n_e}=-1$ and
$\ppx{f}=-\wz\cos\pi x/L_x$ so that $\nefl$ and $\ffl$ are driven
differently regardless of the value of the gradient parameter $\wz$.
The domain size is given by $L_x=L_y=2\pi/K$, with $K$ nominally set to
$0.1$ along with nominal values of $D=0.1$ and $\wz=1$.  There are
$64\times 64$ grid nodes for all cases.  The timestep is set to
$\tau=0.05$.  All runs which reached saturation
were taken to $t=1000$ with
the interval $400<t<1000$ as the saturated stage.

The variations for the drift wave cases with $\wcv=\nu_d=0$ were \hfb
$D=\{0.001,0.003,0.01,0.03,0.1,0.3,1.,3.\}$ with
nominal $K$ and $\wz$, $K=\{0.025,0.05,0.1\}$ at $D=0.01$ and nominal
$\wz$, and $\wz=\{0.1,0.2,0.5,1.\}$ at nominal $D$ and $K$.  The cases
with $D=0.001$ and $0.003$ did not saturate, as thin radial flow jets
formed and simply grew, so another scan with $D=\{0.001,0.003,0.01\}$
was done at $K=0.025$ and nominal $\wz$.  In all cases in which
saturation was found, the
cross correlation between $\phifl$ and $\nefl$
was larger than $0.85$.  This is due to the nature of the turbulence in
this model: at low $D$ the spectrum migrates to large enough scale that
$D$ can compete with $\kkpp\omega$, with $\omega$ close to 
the diamagnetic value of $k_y$, thereby guaranteeing strong adiabatic
coupling for the long-wave side of the energy containing range.  At high
$D$, with all energetic transfer phenomena becoming weak at large scale,
the turbulence again migrated there, and with the weak nonlinearities
the cases for $D>1$ are strongly adiabatic with most of the energy again
at long wavelength; the $D=3.0$ case was taken to $t=4000$ with the
saturation stage taken as the $2000<t<4000$ interval.
All of this has been the subject of previous study
\cite{wakhas,camargo}.  

\begin{figure}[H]
\includegraphics{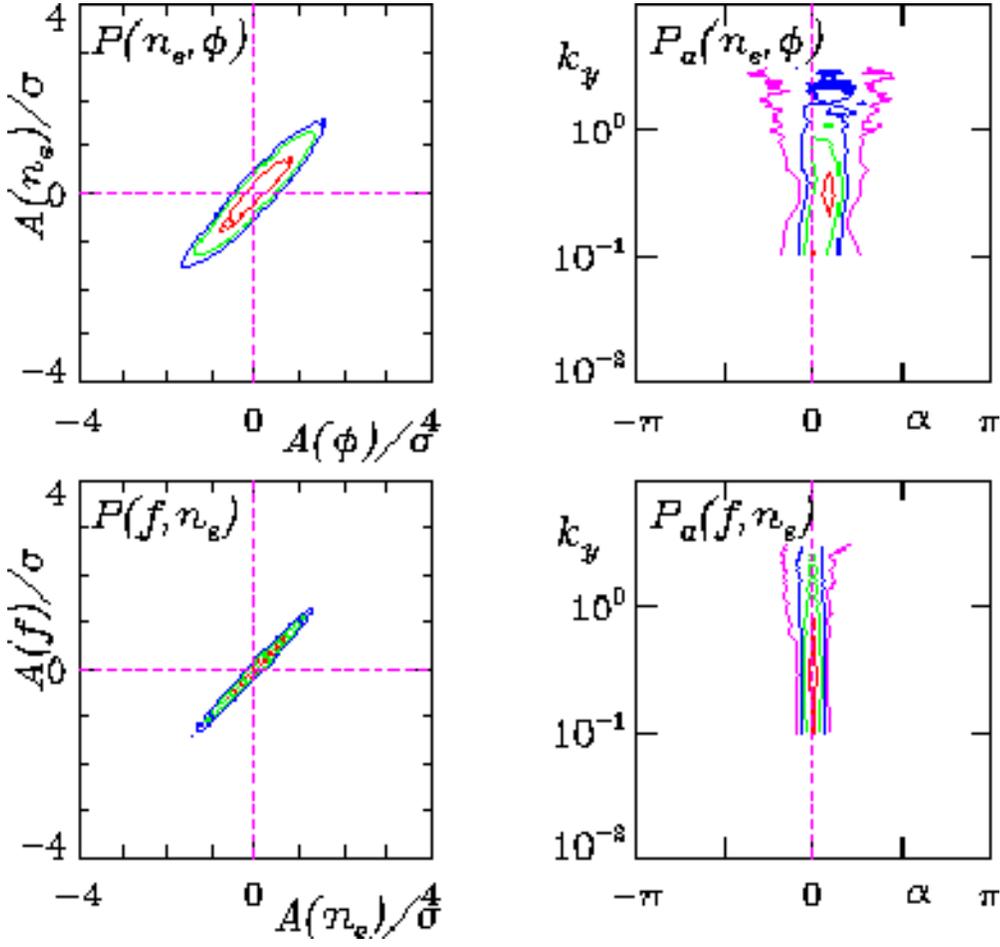}
\caption{
Control check using the nominal case from the 2D dissipative drift wave
model with a test particle density field $\ffl$.
Cross correlation and phase shift distributions,
with $k_y$ in units of $\rho_s^{-1}$, between 
the electron density ($\nefl$) and the electrostatic potential
($\phifl$), 
and between $\ffl$ and $\nefl$.
The first pair shows the drift wave
mode structure and the second pair shows the dynamical test density
alignment.
}
\label{fighwtpa}
\end{figure}

Regardless of any of this behaviour, the main point of this present
study remained.  The cross correlation between $\nefl$ and $\ffl$ was
measured at $0.984$ or larger for all cases.  For $D=0.01$ and larger
at nominal $K$ and $\wz$ these correlation values were \hfb
$\{0.999,0.998,0.992,0.985,0.984,0.987\}$ respectively.  For $\wz=0.1$
and larger at nominal $D$ and $K$ all values were identical at $0.992$.
For $K=0.025$ and larger at $D=0.01$ and nominal $\wz$ they were
$\{0.998,0.9994,0.999\}$.  For $D=0.001$ and larger at 
$K=0.025$ and nominal $\wz$ they were $\{0.99995,0.9998,0.998\}$.
The cross correlation and phase shift distributions between $\nefl$ and
$\phifl$ and between $\ffl$ and $\nefl$ are shown in Fig.\ \ref{fighwtpa}
for the nominal case.
The first pair show the expected near-adiabatic drift wave
mode structure and the second pair show very close correlation between
the densities.

\begin{figure}[H]
\includegraphics{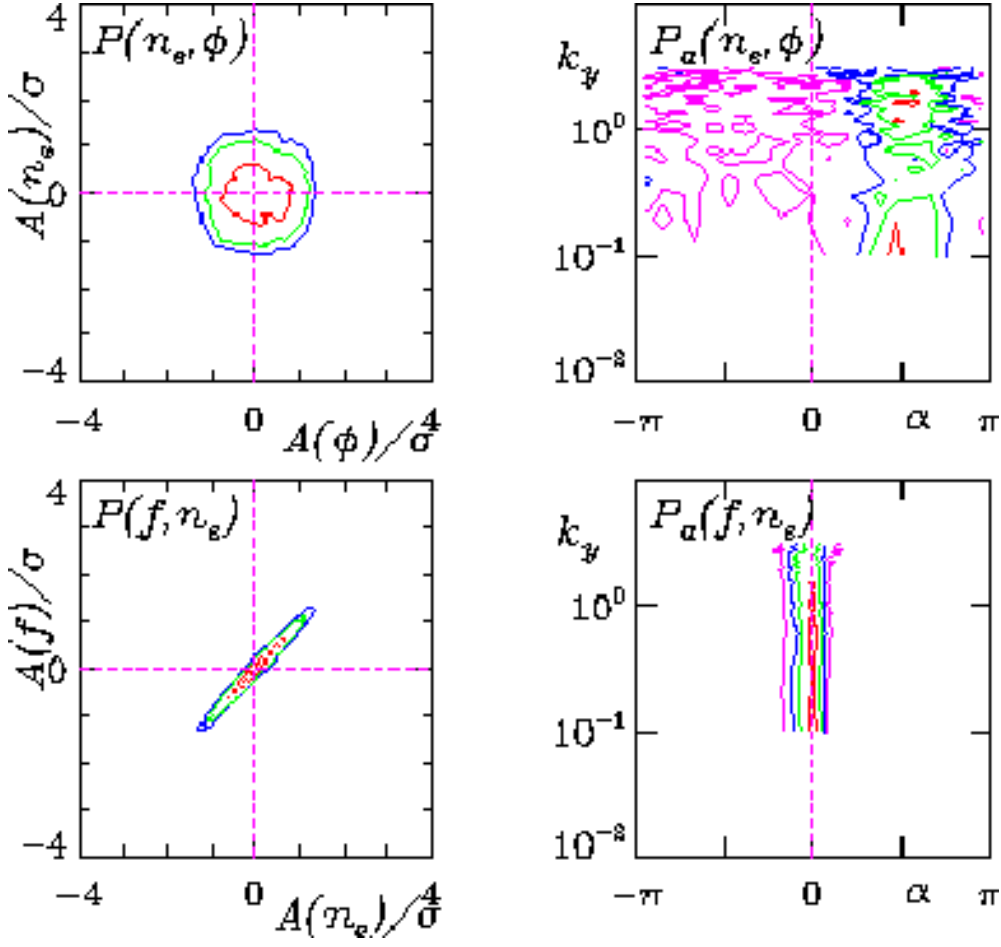}
\caption{
A further control check using the 2D interchange 
model with a test particle density field $\ffl$.
Cross correlation and phase shift distributions,
with $k_y$ in units of $\rho_s^{-1}$, between 
the electron density ($\nefl$) and the electrostatic potential
($\phifl$), 
and between $\ffl$ and $\nefl$.
The first pair shows hydrodynamic
mode structure and the second pair shows the dynamical test density
alignment.
}
\label{fighwtpb}
\end{figure}

The purely hydrodynamic limit is found by setting $D=0$ and then
$\wcv=0.1$ to drive interchange turbulence and $\nu_d=0.01$ to allow it
to saturate.  The cross correlation and phase shift distributions
for this case
between $\nefl$ and $\phifl$ and between $\ffl$ and $\nefl$ are shown in
Fig.\ \ref{fighwtpb}.  The first pair show the expected hydrodynamic
mode structure and the second pair show very close correlation between
the densities.

The phenomena of dynamical alignment
therefore remained for all cases, surviving the control check
represented by this simple 2D drift wave/interchange model.

\section{Conclusions
\label{sec:concl}}

A set of three dimensional computations of turbulence in an
electromagnetic drift wave regime commensurate with tokamak edge
turbulence of a fully self consistent, two-component plasma has been
given.  The principal focus has been on the second ion species as a
proper test species, namely, with no back reaction on the polarisation,
as controlled by a background charge density parameter nominally set to
zero.  Variation of this parameter was found to produce negligible
changes.  Both decaying and source driven cases were considered.  This
difference also produced no difference in turbulence regime, though with
the sources present the turbulence and transport were able to fully
saturate, leaving more activity at very small scale at late times.
A set of two dimensional computations within a dissipative drift wave
model with a test species continuity equation was also given as a
control case.  The numerical schemes and boundary conditions in the 2D
and 3D models were different.  Nevertheless, the same conclusion with
regard to the correlation of the test ion species was always the same:
the test ion and electron densities were found to be correlated
to much better than 80 percent in all cases, and better than 90
percent in most, as measured outside the source region in the 3D driven cases.

It is important to note that the relationship between the electron
density itself and the electrostatic potential (the stream function for
the ExB eddies) was itself quite different among some of these cases.
The adiabatic response of the potential back to the electrons is
parameter dependent, with the electrons usually not in a test particle
relationship vis-a-vis the eddies.  Nevertheless, so
long as there is a significant nonadiabatic electron density response,
the test particle transport was found to conform to whatever the
electron transport result was.

Indeed, these result suggest that the character of the adiabatic
response is not the electrons following the potential, but the other way
around: the potential pushes all species similarly through ExB
advection, but the form of its disturbances and hence the ExB eddies
follows that of the electron pressure disturbances due to the
adiabatic response.  The similarity of transport of all species
regardless of fractional composition now becomes an expected result,
since the ExB advection is the most robust effect in any of the
continuity equations.  The adiabatic response is therefore mostly a
matter of the parallel current mediating the dynamics of polarisation.

There is an important experimental implication in this:  Experiments are
underway to measure the transport of trace tritium ions in tokamaks
\cite{traceexpt}.  Transport of the trace ions is to be compared to that
of the electrons.  It is very important to note that if the two
transports are found to have the same character, it does not follow that
the electrons themselves transport like test particles, namely, that
they should follow a random walk process.  The existence of adiabatic
coupling makes the electron transport different from this.  The results
in this work show that the potential finding that trace ion transport
would be similar to the electron transport does not by itself give an
indication of the character of the electron transport itself.

It remains to establish what these dynamical relationships become in a
regime where the passing electrons are completely adiabatic, and
particle transport is solely due to trapped electrons, and thermal
transport is mostly due to the ions, through the standard ITG scenario
which receives most current attention \cite{dimits}.  Moreover, the
question of an impurity driven particle pinch remains open, since
the actual temperature fluctuation and gradient
dynamics is not included in the GEM3 model.  
Future
manifestations of this work will address these questions with a more
detailed model \cite{gem} designed to treat them in either edge or core
plasma regimes.

\par\vfill\eject

{
\bibliography{paper}
\bibliographystyle{aip}
}

\par\vfill\eject

\end{document}